\documentclass[11pt,superscriptaddress]{article}
\usepackage{jheppub}
\usepackage{epsfig}
\usepackage{amsmath,amssymb}
\usepackage{amsfonts}
\usepackage{mathrsfs}
\usepackage{relsize}
\RequirePackage{xspace}
\usepackage{graphicx}
\usepackage{slashed}
\usepackage{bm}

\title{$\alpha_sv^2$ corrections to $\eta_c$ and  $\chi_{cJ}$ production recoiled with a photon at $e^+e^-$ colliders}

\author[a]{Guang-Zhi Xu }
\emailAdd{ still200@gmail.com}
\author[b]{, Yi-Jie Li }
\emailAdd{ yijiegood@gmail.com}
\author[b]{, Kui-Yong Liu }
\emailAdd{liukuiyong@lnu.edu.cn}
\author[a,c]{, Yu-Jie Zhang}
\emailAdd{nophy0@gmail.com}

\affiliation[a]{School of Physics,
  Beihang University, Beijing 100191, China}

\affiliation[b]{Department of Physics, Liaoning University, Shenyang 110036
, China}

\affiliation[c]{CAS Center for Excellence in Particle Physics, Beijing 100049, China}

\abstract{ We consider the production of the $\eta_c$ and $\chi_{cJ}$ states recoiled with a photon up to $\mathcal{O}(\as v^2)$ at BESIII and B-factories
within the frame of NRQCD factorization.
With the corrections, we revisit the numerical calculations to the cross sections for the $\eta_c(nS)$ and the $\chi_{cJ}(mP)$ states.  We argue that the
search for $XYZ$ states  with even charge conjugation  such as $X(3872)$, $X(3940)$,  $X(4160)$, and $X(4350)$ recoiled with a photon at BESIII may help clarify the nature of
these states.
For completeness, the production of  charmonium  with even charge conjugation
recoiled with a photon at B factories is also discussed.

}

\def \bqn {\begin{eqnarray}}
\def \eqn {\end{eqnarray}}
\def \bqna {\begin{eqnarray}}
\def \eqna {\end{eqnarray}}
\def \eps {\epsilon}
\def \as {\alpha_s}
\def \jpsi {J/\psi}
\def \ee {e^+e^-}

\begin{document}
\maketitle
\flushbottom

\section{Introduction}

Non-relativistic quantum chromodynamics (NRQCD) is a rigorous and successful effect field theory that describes heavy quarkonium decay and production\cite{Bodwin:1994jh}.
The color-octet mechanics (COM) is proposed in NRQCD.
The infrared divergences in the decay
widths of $P$-wave \cite{Brambilla:2008zg,Lansberg:2009xh} and $D$-wave \cite{He:2008xb,He:2009bf,Fan:2009cj}  heavy quarkonium have been absorbed into the NRQCD matrix elements applied with COM, and the infrared-safe decay rate can be obtained.
But the last decade experiment measurements at $\ee$ colliders and at hadron colliders reveal large discrepancies with LO (leading order) calculations.

In the $\ee$ annihilation experiment\cite{Abe:2002rb,Aubert:2005tj}, problems on NRQCD involving the inclusive and exclusive $\jpsi$ production\cite{Liu:2002wq,Braaten:2002fi,Liu:2003jj,Liu:2003zr,Liu:2004ga,Liu:2004un} had been solved by higher-order corrections, including radiative corrections\cite{Zhang:2005cha,Zhang:2006ay,Wang:2011qg,Zhang:2008gp,Zhang:2009ym,
Gong:2007db,Gong:2008ce,Gong:2009ng,Gong:2009kp,Ma:2008gq,Dong:2011fb,
Bodwin:2014dqa}, relativistic corrections\cite{He:2007te,Bodwin:2006ke,Ebert:2006xq,
Bodwin:2007ga,Elekina:2009wt,Jia:2009np,He:2009uf,Fan:2012dy,Fan:2012vw}, and ${\mathcal O}(\alpha_s v^2)$ corrections \cite{Dong:2012xx,Li:2013qp}.
And the LO NRQCD calculations at hadron colliders also encounters dilemma in the heavy quarkonium  production  and polarization especially at the large $p_t$ region.
The contributions from the NLO (next-to-leading order) radiative corrections to the heavy quarkonium  production
\cite{Campbell:2007ws,Gong:2008hk,Gong:2008sn,Gang:2012js,Ma:2010yw,
Ma:2010vd,Ma:2010jj,Shao:2012iz,Butenschoen:2010rq,Wang:2012is,Butenschoen:2013pxa,
Meng:2013gga,Li:2014ava,Wang:2014vsa,Sun:2014gca,Bodwin:2014gia}
and polarization \cite{Chao:2012iv,Butenschoen:2012px,Gong:2012ug,Gong:2013qka,Shao:2012fs,Shao:2014fca}
at hadron colliders are significant. And the contributions of the NLO relativistic corrections to
$J/\psi$ hadronic production are considered too\cite{Fan:2009zq,Xu:2012am,Li:2013csa}. $O(\alpha_{s}v^2)$ corrections to
the decays of $h_c, h_b$ and $\eta_b$ are studied in Ref.\cite{Jia:2011ah,Guo:2011tz,Li:2012rn}.
Actually, the corrections at higher-order(e.g., $\mathcal{O}(\as v^2)$, $v^4$), had been considered in many processes and contributed considerable effects.
However, some drawbacks for fixed-order calculations involve the convergence for higher-order corrections and to which order should be considered within NRQCD.
These problems can be understood by adding more higher-order calculations. More information about NRQCD can be found in Ref.\cite{Brambilla:2010cs} and related papers.

Studies have focused on the production of quarkonium with even charge conjugation that are recoiled with a hard photon in the $\ee$ annihilation at the B factories and BESIII is a very interesting process. The production of
double charmonium at
B factories\cite{Abe:2002rb,Aubert:2005tj}  aids in identifying to identify
charmonium or charmonium-like states with even charge conjugation, which recoiling
with $J/\psi$ and
$\psi(2S)$.  $\eta_c, \eta_c(2S)$, $\chi_{c0}$,
 $X(3940)$ (decaying into $D\bar {D^*}$), and $X(4160)$ (decaying
into $D^*\bar {D^*}$) have also been observed in double charmonium
production at B factories, but the  $\chi_{c1}$ and $\chi_{c2}$ states
are yet to be determined in production associated with $J/\psi$ at B factories.
 The LO  calculation for  heavy quarkonium  with even charge conjugation  recoiled with a hard photon in the $\ee$ annihilation at the B factories and BESIII  is a pure QED process\cite{Chung:2008km,Braguta:2010mf}.
The one-loop calculations have been computated and analyzed\cite{Li:2009ki,Sang:2009jc,Sang:2012cp,Li:2013nna,Chao:2013cca}. And the NLO relativistic corrections have been computed too \cite{Sang:2009jc,Li:2013nna}.
Quarkonium with even charge conjugation are associated to the $XYZ$ particles\cite{Li:2009zu,Molina:2009ct,Li:2012vc}.
The well-known one of the $XYZ$ particles, $X(3872)$\cite{Choi:2003ue}, is supposed to the $\chi_{c1}^{\prime}$ state or the mixture of this state with other structure in some view\cite{Meng:2005er,Meng:2013gga}.
Recently, $X(3872)$ has been observed in photon-recoiled process with a statistical significance of $6.4\sigma$ at BESIII\cite{Ablikim:2013dyn}.
 $X(3915)$ ($X(3945)$ or $Y(3940)$) and $Z(3930)$ are assigned as the
$\chi_{c0}(2P)$ and $\chi_{c2}(2P)$ states by the PDG (Particle Data Group)\cite{Beringer:1900zz}. However this identification may be called into some questions\cite{Guo:2012tv}.
The experimental results for states with even charge conjugation have theoretically elicited interest in the nature of charmonium-like states. The non-perturbative effects are strong because the energy region at BESIII approximates the threshold charmonium states. Hence, the applicability of NRQCD is speculative within this region. However, some NRQCD-based calculations exhibit high compatibility with the data.

In this paper, the photon-recoiled $\eta_c$  and $\chi_{cJ}$ production is studied according to our previous work\cite{Li:2013nna}. We calculate the cross sections up to the order of $\mathcal{O}(\as v^2)$ within the NRQCD. This study verifies the applicability of NRQCD at the threshold and determines the $XYZ$ particles related to $\eta_c(nS)$ and $\chi_{cJ}(nP)$.

The paper is organized as follows. Sec.\ref{sec:frame} introduces  the framework of calculations, especially the method of the expansion up to $\mathcal(\as v^2)$ for the amplitudes. Sec.\ref{sec_cs} presents the amplitudes expansion and discussion the cross sections for the $\eta_c$ and $\chi_{cJ}$ process. Sec.\ref{sec:numer} gives the numerical results up to $\mathcal{O}(\as v^2)$. Finally, Sec.\ref{sec_summary} presents a summary.

\section{The framework of the calculation}\label{sec:frame}

This section introduces the calculation method for
the $\mathcal{O}(\as v^2)$ amplitude expansion to the process $e^+e^-\to \gamma^\ast \to H (\eta_c, \chi_{cJ}) + \gamma$. The momenta of final states are stated as $H (p) $~ and ~$\gamma (k)$.
Cross section can be obtained and applied to express the amplitudes via expansions.

\subsection{Kinematics}\label{s_kinematics}

In an arbitrary fame of the charmonium, the momenta of the charm and the anti-charm can be expressed by the meson momentum and their relative momentum,
\begin{eqnarray}\label{q_pcexd}
p_c=p/2+q,\nonumber\\
p_{\bar{c}}=p/2-q.
\end{eqnarray}
The momenta $p$ and $q$ are orthogonal, i.e.,  $p{\cdot}q=0$. In the meson rest frame, they can be written as, $p=(2E_q,\textbf{0})$ and $q=(0,\textbf{q})$.
We calculated the amplitudes up to the order $\mathcal{O}(\as v^2)$ using an orthodox method. In this method, the rest energy $E_q=\sqrt{m_c^2+\textbf{q}^2}$ of the charm/anti-charm should be expanded around the charm mass,
\begin{eqnarray}\label{q_eqexd}
E_q=m_c+\frac{\textbf{q}^2}{m_c^2}\frac{m_c}{2}+\mathcal{O}(\frac{\textbf{q}^4}{m_c^4}).
\end{eqnarray}

The momenta of the final-state particles depend on $E_q$. For instance, the four-momenta of the particles in $ \gamma^\ast (Q)\to H (p) + \gamma (k)$ in the center-of-mass system  can be written as follows:
\begin{eqnarray}\label{momINCOM}
Q&=&(\sqrt{s},0,0,0),\nonumber\\
p&=&(\frac{s+4E_q^2}{2\sqrt s},0,0,\frac{s-4E_q^2}{2\sqrt s}),\nonumber\\
k&=&(\frac{s-4E_q^2}{2\sqrt s},0,0,-\frac{s-4E_q^2}{2\sqrt s}).
\end{eqnarray}

Given the expression for $E_q$, the four-momenta can be expanded  in terms of $\textbf{q}^2/m_c^2$.
For instance, the momenta of the final meson and the photon noted by $p$ and $k$ are expanded  as the following expression,
\begin{eqnarray}\label{q_momexd}
p&=&(\frac{s+4m_c^2}{2\sqrt s},0,0,\frac{s-4m_c^2}{2\sqrt s})+\frac{\textbf{q}^2}{m_c^2}\frac{2m_c^2}{\sqrt s}(1,0,0,-1)+\mathcal{O}(\frac{\textbf{q}^4}{m_c^4})\nonumber\\
&=&p^{(0)}+\frac{\textbf{q}^2}{m_c^2}p^{(2)}+\mathcal{O}(\frac{\textbf{q}^4}{m_c^4}),\nonumber\\
k&=&\frac{s-4m_c^2}{2\sqrt s}(1,0,0,-1)-\frac{\textbf{q}^2}{m_c^2}\frac{2m_c^2}{\sqrt s}(1,0,0,-1)+\mathcal{O}(\frac{\textbf{q}^4}{m_c^4})\nonumber\\
&=&k^{(0)}+\frac{\textbf{q}^2}{m_c^2}k^{(2)}+\mathcal{O}(\frac{\textbf{q}^4}{m_c^4}).
\end{eqnarray}
Therefore, the momenta with subscripts $(0)$ or $(2)$ are independent of $\textbf{q}^2$.
The scalar products of ($p^{(0)}$, $p^{(2)}$, $k^{0}$,  and $k^{(2)}$) can be solved in a special frame.
For instance, in the center-of-mass system, the relation $k^{(2)}=-p^{(2)}$ can be obtained to reduce the number of the independent momenta; all the three non-zero products are calculated as follows:
\begin{eqnarray}\label{q_sp}
p^{(0)}{\cdot}p^{(0)}&=&4m_c^2, \nonumber \\
 p^{(0)}{\cdot}p^{(2)}&=&2m_c^2, \nonumber \\
  p^{(0)}{\cdot}k^{(0)}&=&(s-4m_c^2)/2.
\end{eqnarray}

Studies on the $\mathcal{O}(\as v^2)$ corrections to the decay process of charmonium with massless final-states(\cite{Bodwin:2002hg,Li:2012rn,Jia:2011ah,Guo:2011tz}) introduce a factor $E_q/m_c$ to all external momenta.
In our method, these momenta can be expanded as $p_i=p_i^{(0)}+\frac{\textbf{q}^2}{m^2}p_i^{(2)}=p_i^{(0)}(1+\frac{\textbf{q}^2}{2m_c^2})$ with $p_i^{(2)}=p_i^{(0)}/2$. This equation indicates the compatibility of our method with that published.

For the $P$-wave states, the spin and orbital vectors must also be expanded by
\begin{eqnarray}
&&\eps_s=\eps_s^{(0)}+\frac{\textbf{q}^2}{m_c^2}\eps_s^{(2)}+\mathcal{O}(\frac{\textbf{q}^4}{m_c^4}),\nonumber\\
&&\eps_L=\eps_L^{(0)}+\frac{\textbf{q}^2}{m_c^2}\eps_L^{(2)}+\mathcal{O}(\frac{\textbf{q}^4}{m_c^4}),
\end{eqnarray}
Furthermore, they couple onto the total angular momentum $J$ states ($J=0,1,2$)  with the relation presented as follows:
\begin{eqnarray}
&&\mathcal{P}_0^{\alpha\beta}\equiv\sum_{s_zL_z}\eps_s^{\ast\alpha}
\eps_L^{\ast\beta}\langle1s_z;1L_z|00\rangle=\frac{1}{\sqrt{D-1}}\Pi^{\alpha\beta},\nonumber\\
&&\mathcal{P}_1^{\alpha\beta}\equiv\sum_{s_zL_z}\eps_s^{\ast\alpha}
\eps_L^{\ast\beta}\langle1s_z;1L_z|1J_z\rangle=\frac{i}{\sqrt{2}M}
\eps^{\alpha\beta\kappa\lambda}p_{\kappa}\eps_{\lambda}^{\ast}(J_z),\nonumber\\
&&\mathcal{P}_2^{\alpha\beta}\equiv\sum_{s_zL_z}\eps_s^{\ast\alpha}
\eps_L^{\ast\beta}\langle1s_z;1L_z|2J_z\rangle=\eps^{\ast\alpha\beta}(J_z).
\end{eqnarray}
The polarization is summed over all directions of the vector for the total angular momentum:
\begin{eqnarray}
&&\sum_{J_z}\eps^{\alpha}(J_z)\eps^{\ast\beta}(J_z)=\Pi^{\alpha\beta},\nonumber\\
&&\sum_{J_z}\eps^{\alpha\beta}\eps^{\ast\alpha^\prime\beta^\prime}=
\frac{1}{2}(\Pi^{\alpha\alpha^\prime}\Pi^{\beta\beta^\prime}+
\Pi^{\alpha\beta^\prime}\Pi^{\alpha^\prime\beta})
-\frac{1}{D-1}\Pi^{\alpha\beta}\Pi^{\alpha^\prime\beta^\prime},
\end{eqnarray}
where $\Pi$ can be expanded in terms of  $\textbf{q}$:
\begin{eqnarray}\label{eq_piexd}
&&\Pi_{\alpha\beta}\equiv-g_{\alpha\beta}+\frac{p_{\alpha}p_{\beta}}{p^2},\nonumber\\
&&\Pi_{\alpha\beta}=\Pi^{(0)}_{\alpha\beta}+
\frac{\textbf{q}^2}{4m_c^2}(p^{(0)}_{\alpha}p^{(2)}_{\beta}
+p^{(2)}_{\alpha}p^{(0)}_{\beta}-p^{(0)}_{\alpha}p^{(0)}_{\beta})+\mathcal{O}(\frac{\textbf{q}^4}{m_c^4}).
\end{eqnarray}
The second term vanishes in the rest frame of the meson, which is consistent with the independence of the polarization vectors to $\textbf{q}^2$ in this frame.

\subsection{Amplitudes expansion}

The amplitude of $e^+e^-{\rightarrow}\gamma H(\eta_c, \chi_{cJ})$ can be written as\cite{Bodwin:2007ga}
\begin{eqnarray}
\mathcal{M}(e^+e^-{\rightarrow}\gamma H)=L_{\alpha}\mathcal{M}^{\alpha}(\gamma^\ast\rightarrow\gamma H),
\end{eqnarray}
where the leptonic part $L_{\alpha}$ is independent of $\textbf{q}$.
We only consider the hadronic part element $\mathcal{M}^{\alpha}(\gamma^\ast\rightarrow\gamma H)$ in the NRQCD frame. The Feynman diagrams are shown in Fig.\ref{fig:feyndia}.
The amplitude can be written as\cite{Bodwin:2007ga}:
\begin{eqnarray}
\mathcal{M}(\gamma^\ast\rightarrow\gamma H)=\sqrt{2M_{H}}\sum_n{d_n\langle H|\mathcal{O}_n^{H}|0\rangle},
\end{eqnarray}
where the factor $\sqrt{2M_{H}}$ originates from the relativistic normalization.
$d_n$ is the short-distance coefficient that can be obtained by
matching with the full QCD calculations on the intermediate $c\bar{c}$ production.
And the $\langle H|\mathcal{O}_n^{H}|0\rangle$ represents the NRQCD long-distance matrix
elements that are extracted from the experimental data or determined by potential model or
lattice calculations. The present study  concentrates on the corrections up to the order
$\mathcal{O}(\as v^2)$ under the color-singlet frame. The expansion is given as follows:
\begin{eqnarray}\label{eq_nrqcd_amp_ex_asv2}
&&\hspace{-1.5cm}\mathcal{M}(\gamma^\ast\rightarrow\gamma H)\nonumber \\
&=&\sqrt{2M_{H}}\big[(d^{(0)}+d^{(\as)})\langle H|\mathcal{O}^{H}|0\rangle+(d^{(v^2)}+d^{(\as v^2)})\langle H|\mathcal{P}^{H}|0\rangle\big]\nonumber\\
&\approx&2\sqrt{m_c}(1+\frac{\textbf{q}^2}{4m_c^2})\big[(d^{(0)}+d^{(\as)})\langle
H|\mathcal{O}^{H}|0\rangle+(d^{(v^2)}+d^{(\as v^2)})\langle H|\mathcal{P}^{H}|0\rangle\big].
\end{eqnarray}

The short-distance coefficients are obtained from the matching between
the pQCD  and the NRQCD calculations on the $c\bar{c}$ production,
\begin{eqnarray}\label{eq_matching}
&&\hspace{-1.5cm}\mathcal{M}_{s}\big[\gamma^\ast\rightarrow\gamma+c\bar{c}\big]|_{pQCD}\nonumber \\&=&(d^{(0)}_s+d^{(\as)}_s)\langle c\bar{c}
|\mathcal{O}^{c\bar{c}}(^1S_0^{[1]})|0\rangle+(d^{(v^2)}_s+d^{(\as v^2)}_s)\langle c\bar{c}|\mathcal{P}^{c\bar{c}}(^1S_0^{[1]})|0\rangle
\nonumber\\
&=&\sqrt{2N_c}2E_q\big[(d^{(0)}_s+d^{(\as)}_s)+\textbf{q}^2(d^{(v^2)}_s+d^{(\as v^2)}_s+d^{(self.)}_s)\big].\nonumber\\
&&\hspace{-1.5cm}\mathcal{M}_{t}\big[\gamma^\ast\rightarrow\gamma+c\bar{c}\big]|_{pQCD}\nonumber \\&=&(d^{(0)}_{t}+d^{(\as)}_{t})\langle c\bar{c}|\mathcal{O}^{c\bar{c}}(^3P_J^{[1]})|0\rangle+(d^{(v^2)}_{t}+d^{(\as v^2)}_{t})\langle c\bar{c}|\mathcal{P}^{c\bar{c}}(^3P_J^{[1]})|0\rangle
\nonumber\\
&=&\sqrt{2N_c}2E_q\big[|\textbf{q}|(d^{(0)}_{t}+d^{(\as)}_{t})+\textbf{q}^3(d^{(v^2)}_{t}+d^{(\as v^2)}_{t}+d^{(self.)}_{t})\big],
\end{eqnarray}
where $\mathcal{M}_{s}$ and $\mathcal{M}_{t}$ represent the amplitudes with
the $c\bar{c}$ pair coupling to spin-singlet and spin-triplet polarization, respectively.
The above NRQCD operators $\mathcal{O}$ and  $\mathcal{P}$ are respectively defined as follows:
\begin{eqnarray}
&&\mathcal{O}^{c\bar{c}}(^1S_0^{[1]})=\psi^\dagger\chi,\,\nonumber\\
&&\mathcal{P}^{c\bar{c}}(^1S_0^{[1]})=\psi^\dagger(-\frac{i}{2}
\overleftrightarrow{\textbf{D}})^2\chi,\nonumber\\
&&\mathcal{O}^{c\bar{c}}(^3P_0^{[1]})=\frac{1}{3}\psi^\dagger(\frac{-i}{2}
\overleftrightarrow{\textbf{D}}{\cdot}{\sigma})\chi,\nonumber\\
&&\mathcal{P}^{c\bar{c}}(^3P_0^{[1]})=\frac{1}{3}\psi^\dagger[(-\frac{i}{2}
\overleftrightarrow{\textbf{D}})^2(\frac{-i}{2}\overleftrightarrow{\textbf{D}}{\cdot}{\sigma})]\chi,\nonumber\\
&&\mathcal{O}^{c\bar{c}}(^3P_1^{[1]})=\frac{1}{2}\psi^\dagger(\frac{-i}{2}
\overleftrightarrow{\textbf{D}}{\times}{\sigma})\chi,\nonumber\\
&&\mathcal{P}^{c\bar{c}}(^3P_1^{[1]})=\frac{1}{2}\psi^\dagger[(-\frac{i}{2}
\overleftrightarrow{\textbf{D}})^2(\frac{-i}{2}\overleftrightarrow{\textbf{D}}{\times}{\sigma})]\chi,\nonumber\\
&&\mathcal{O}^{c\bar{c}}(^3P_2^{[1]})=\psi^\dagger(\frac{-i}{2}
\overleftrightarrow{D}^{(i}{\sigma}^{j)})\chi,\nonumber\\
&&\mathcal{P}^{c\bar{c}}(^3P_2^{[1]})=\psi^\dagger[(-\frac{i}{2}
\overleftrightarrow{\textbf{D}})^2(\frac{-i}{2}
\overleftrightarrow{D}^{(i}{\sigma}^{j)})]\chi,
\end{eqnarray}
where Pauli spinors $\psi$ and $\chi$ describe the quark
annihilation and the anti-quark creation, respectively. The gauge-covariant derivative operator $\overleftrightarrow{\textbf{D}}=\overrightarrow{\textbf{D}}-\overleftarrow{\textbf{D}}$.
The term $d^{(self.)}$ originates from the one-loop self-energy corrections to the NRQCD matrix elements\cite{Bodwin:1994jh,Dong:2012xx,Guo:2011tz,Jia:2011ah} and in the $\overline{MS}$ scheme
\begin{eqnarray}\label{eq:MErenormalization}
\langle c\bar{c}|\mathcal{O}^{c\bar{c}}|0\rangle_{\overline{MS}}=(\langle c\bar{c}|\mathcal{O}^{c\bar{c}}|0\rangle)^{(0)}+\frac{2\as}{3{\pi}m_Q^2}C_F\frac{N_{\eps}}{\eps_{IR}}(\langle c\bar{c}|\mathcal{P}^{c\bar{c}}|0\rangle)^{(0)},
\end{eqnarray}
where $N_{\epsilon}(m_Q)\equiv(\frac{4\pi{\mu}_r^2}{m_Q^2})^{\epsilon}\Gamma(1+\epsilon)$.
${\mu}_r$ is the renormalization scale.
Therefore,
\begin{eqnarray}\label{eq:dselfMErenormalization}
d^{(self.)}=\frac{2\as}{3{\pi}m_Q^2}C_F[\frac{1}{\epsilon_{IR}}
+\ln4\pi-\gamma_E+\ln(\frac{\mu_r^2}{m_Q^2})]d^{(0)}.
\end{eqnarray}
This expression is  satisfied for all $^1S_0^{[1]}$ and $^3P_J^{[1]}$ states.
Therefore, $d^{(self.)}$ contributes to the amplitudes expansion for $\mathcal{O}(\as v^2)$.
The factor $\sqrt{2N_c}2E_q$ in Eq.(\ref{eq_matching}) originates from
the perturbative calculations on the LO $Q\overline{Q}$ NRQCD matrix elements.
The extra factor $|\textbf{q}|$ arises from the derivative operator for the $P$-wave NRQCD operator $\mathcal{P}$.

The covariant projection method is adopted to calculate the full QCD amplitudes as,
\begin{eqnarray}
&&\mathcal{M}_s\big[\gamma^\ast\rightarrow\gamma+c\bar{c}\big]
=Tr\{\mathcal{M}[\gamma^\ast\rightarrow\gamma+c+\bar{c}){\otimes}\mathcal{P}_{00}{\otimes}\pi_1]\},\nonumber\\
&&\mathcal{M}_t\big[\gamma^\ast\rightarrow\gamma+c\bar{c}\big]
=Tr\{\mathcal{M}[\gamma^\ast\rightarrow\gamma+c+\bar{c}){\otimes}\mathcal{P}_{1s_z}{\otimes}\pi_1]\},
\end{eqnarray}
The color-singlet projection operators are defined as $\pi_1=\textbf{1}/\sqrt{N_c}$. The spin-singlet and spin-triplet projection operators are given as,
\begin{eqnarray}\label{q_proj}
&&\mathcal{P}_{00}=\frac{1}{2\sqrt{2}(E_q+m_c)}(\slashed{p}_{\bar{c}}-m_c)
\frac{(-\slashed{p}+2E_q)\gamma_5(\slashed{p}+2E_q)}{8E_q^2}(\slashed{p}_c+m_c),\nonumber\\
&&\mathcal{P}_{1s_z}(\eps_s)=\frac{1}{2\sqrt{2}(E_q+m_c)}
(\slashed{p}_{\bar{c}}-m_c)\frac{(-\slashed{p}+2E_q)\slashed{\eps}_s(\slashed{p}+2E_q)}{8E_q^2}(\slashed{p}_c+m_c),
\end{eqnarray}
where $\mathcal{P}_{00}$ and $\mathcal{P}_{1s_z}$ for the spin-singlet and spin-triplet states, respectively.
These operators can be expanded up to the $\textbf{q}^2/m_c^2$ order applied with Eqs:(\ref{q_pcexd}),(\ref{q_eqexd}),(\ref{q_momexd}).

According to the matching expression Eq.(\ref{eq_matching}), the short-distance coefficients are calculated by
\begin{eqnarray}
&&d^{(0)}_s=\frac{\mathcal{M}^{(0)}_s}{\sqrt{2N_c}2m_c}\big|_{q\rightarrow0},\nonumber\\
&&d^{(\as)}_s=\frac{\mathcal{M}^{(\as)}_s}{\sqrt{2N_c}2m_c}\big|_{q\rightarrow0},\nonumber\\
&&d^{(v^2)}_s=\frac{1}{2!}\frac{\partial^2}{\partial\textbf{q}^2}
\frac{\mathcal{M}^{(0)}_s}{\sqrt{2N_c}2E_q}\big|_{q\rightarrow0},\nonumber\\
&&d^{(\as v^2)}_s=\frac{1}{2!}\frac{\partial^2}{\partial\textbf{q}^2}
\frac{\mathcal{M}^{(\as)}_s}{\sqrt{2N_c}2E_q}-d^{self.}_s\big|_{q\rightarrow0},\nonumber\\
&&d^{(0)}_t=\eps_L^{(0)}\frac{\partial}{\partial|\textbf{q}|}
\frac{\mathcal{M}^{(0)}_t}{\sqrt{2N_c}2E_q}\big|_{q\rightarrow0},\nonumber\\
&&d^{(\as)}_t=\eps_L^{(0)}\frac{\partial}{\partial|\textbf{q}|}
\frac{\mathcal{M}^{(\as)}_t}{\sqrt{2N_c}2E_q}\big|_{q\rightarrow0},\nonumber\\
&&d^{(v^2)}_t=\eps_L^{(0)}\frac{1}{3!}\frac{\partial^3}{\partial\textbf{q}^3}
\frac{\mathcal{M}^{(0)}_t}{\sqrt{2N_c}2E_q}+\eps_L^{(2)}\frac{d^{(0)}_t}{m_c^2}\big|_{q\rightarrow0},\nonumber\\
&&d^{(\as v^2)}_t=\eps_L^{(0)}\frac{1}{3!}\frac{\partial^3}{\partial\textbf{q}^3}
\frac{\mathcal{M}^{(\as)}_t}{\sqrt{2N_c}2E_q}+\eps_L^{(2)}\frac{d^{(\as)}_t}{m_c^2}-d^{self.}_t\big|_{q\rightarrow0},
\end{eqnarray}
where $\mathcal{M}^{(0)}$ and $\mathcal{M}^{(\as)}$ are defined by the born
and one-loop amplitudes, respectively.
The replacements are applied to resolve the expansion to the Lorentz vector $q$ in the amplitude expressions:
\begin{eqnarray}
q_{\mu}q_{\nu}\rightarrow\frac{\textbf{q}^2}{D-1}\Pi^{(0)}_{\mu\nu},
\end{eqnarray}
for $S$-wave states and
\begin{eqnarray}\label{q_qrep_p}
&&q_{\mu}q_{\nu}q_{\rho}\rightarrow
\frac{\textbf{q}^3}{D+1}\big\{\Pi^{(0)}_{\mu\nu}[\eps_L^{(0)}]_{\rho}
+\Pi^{(0)}_{\mu\rho}[\eps_L^{(0)}]_{\mu}+\Pi^{(0)}_{\nu\rho}[\eps_L^{(0)}]_{\mu}\big\},
\end{eqnarray}
for $P$-wave states.

\subsection{One-loop computation}

\begin{figure}[ht]
\begin{center}
\includegraphics[width=0.98\textwidth]{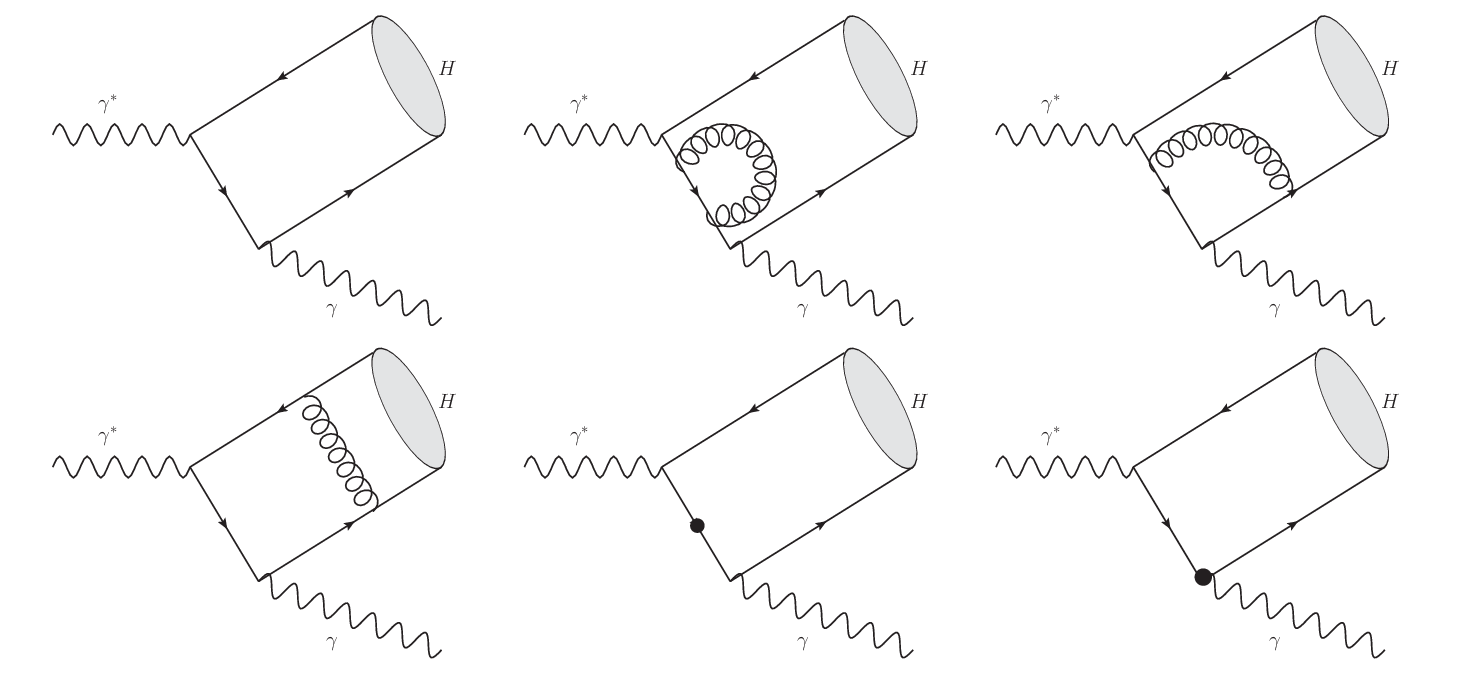}
\end{center}
\caption{\label{fig:feyndia}  The typical born, loop,  and counterterm  Feynman diagrams. There are two diagrams for the born amplitude,  six diagrams for the counterterm amplitude, 
and eight for the one-loop amplitude including two self-energy  diagrams, four triangle  diagrams, and two box diagrams.}
\end{figure}

The one-loop Feynman diagrams are shown in Fig.\ref{fig:feyndia}.
The dimensional regularization scheme is selected here. The ultraviolet divergences in one-loop amplitude are canceled
by the counterterms.
The infrared divergences  at the $\as$ order in one-loop amplitude  are also canceled by the counterterm amplitude, and the additional infrared divergences  at the order of $\as v^2$
 are
canceled by the one-loop self-energy contribution to the NRQCD matrix elements in Eq.\ref{eq:MErenormalization} and Eq.\ref{eq:dselfMErenormalization}.
The real corrections need not to be included for the exclusive processes.
We apply the method in Ref.\cite{Denner:2005nn} to reduce the tensor integration.
The relativistic expansion is done before dealing with the loop integrand.
The on-mass-shell(OS) renormalization scheme is adopted and in
this scheme the renormalization constants are chosen as
\begin{eqnarray}
{\delta}Z_2^{OS}&=&-C_F\frac{\as}{4\pi}N_{\epsilon}\big(\frac{1}{\epsilon_{UV}}+\frac{2}{\epsilon_{IR}}+4\big),\nonumber\\
{\delta}Z_{m_Q}^{OS}&=&-C_F\frac{\as}{4\pi}N_{\epsilon}\big(\frac{3}{\epsilon_{UV}}+4\big),
\end{eqnarray}
where $N_{\epsilon}(m_Q)$ has been previously defined and the renormalization scale ${\mu}_r$
is canceled by the loop and counterterm diagrams up to the order of $\mathcal{O}(\as v^2)$.
In the OS scheme, the diagrams for the external leg correction are not included.
In our calculations, the 't Hooft-Veltman (HV) regularization scheme\cite{'tHooft:1972fi,Breitenlohner:1977hr}  is adopted in which $\gamma^5$ is defined as
\begin{eqnarray}\label{Eq_ga5def}
\gamma^5 {\equiv} i\gamma^0\gamma^1\gamma^2\gamma^3=-\frac{i}{4!}\eps^{\mu\nu\rho\sigma}{\gamma_\mu}{\gamma_\nu}{\gamma_\rho}{\gamma_\sigma},
\end{eqnarray}
The traces involving more than four Dirac $\gamma$-matrices with a $\gamma^5$ are evaluated recursively by the West Mathematica programs\cite{West:1991xv}.
Our strategy handing of $\gamma^5$ is same as that in Ref.\cite{Jia:2011ah}.
In the HV scheme, the Ward identities may be violated in the one-loop calculations, such as for the axial current known as the Adler-Bell-Jackiw anomalies, which arising from the symmetry breaking of $\gamma_5$ definitions in D-dimension as Eq.\ref{Eq_ga5def} . 
In our case,  $\gamma\ast\to\gamma\eta_c$ process,  for $\gamma_5$ appears outside of one-loop integrals,  the amplitudes would satisfy the ward identities, that is seen as the short-distance results given in Eq.\ref{eq_etacsd}  in the next section. 
More discussions of $\gamma^5$-scheme and the anomalous Ward identities could be refered to Refs.\cite{'tHooft:1972fi,'tHooft:1976up,Breitenlohner:1977hr,Larin:1991tj,Korner:1991sx,Kreimer:1993bh,Larin:1993tq,West:1991xv,Jia:2011ah}.

We use the {\tt
FeynArts}\cite{Hahn:2000kx} package to generate Feynman diagrams and amplitudes,
and the {\tt
FeynCalc}\cite{Hahn:1998yk,Mertig:1990an} package and our self-written {\tt
Mathematica} package to handle the amplitudes  and the phase space integrand.

\subsection{Matching results for $\eta_c$}

This subsection presents the matching results for the short-distance coefficients for $\eta_c$.

The final matching results of the coefficients are given in the Appendix,
where  $r\equiv4m_c^2/s$ and $s$ is the square of beam energy.
The coefficients are given as follows
\footnote{Here, we omit the third term of the coefficients in the orders
of $v^2$ and $\as v^2$ as shown in the appendix to dilute the contribution
of the relativistic renormalization in Eq.(\ref{eq_nrqcd_amp_ex_asv2})\cite{Bodwin:2007ga}. In other words, the below coefficients include the contributions of the relativistic renormalization.}:
\begin{eqnarray}\label{eq_etacsd}
d^{(0)}_s&=&A^{(0)}\eps_1,\nonumber\\
d^{(v^2)}_s&=&A^{(v^2)}\eps_1+A^{(0)}\eps_2/m_c^2,\nonumber\\
d^{(\as)}_s&=&A^{(\as)}\eps_1,\nonumber\\
d^{(\as v^2)}_s&=&A^{(\as v^2)}\eps_1+A^{(\as)}\eps_2/m_c^2,
\end{eqnarray}
where
\begin{eqnarray}\label{eq_e1e2def}
&&\eps_1\equiv\eps^{\mu\nu\rho\tau}(\eps^{\ast}_{Q})_{\mu}
(\eps^{\ast}_{k})_{\nu}k^{(0)}_{\rho}p^{(0)}_{\tau},\nonumber\\
&&\eps_2\equiv\eps_1^{(2)}=\eps^{\mu\nu\rho\tau}(\eps^{\ast}_{Q})_{\mu}
(\eps^{\ast}_{k})_{\nu}(p^{(0)}+k^{(0)})_{\rho}p^{(2)}_{\tau},
\end{eqnarray}
where $\eps_Q$ and $\eps_k$ represent the polarization vector of the initial virtual photon and the final photon, respectively.

The coefficients in Eq.(\ref{eq_etacsd}) are provided in the high-energy region.
In the limit $r\rightarrow0$, the asymptotic behavior of these  coefficients can be obtained. The lowest order of the coefficients is  $\mathcal{O}(r)$; the higher-order contributions are omitted, and the reduced equations are given as follows:
\begin{eqnarray}
d^{(0)}_s&=&C\,r\eps_1,\nonumber\\
d^{(v^2)}_s&=&-\frac{5}{12m_c^2}C\,r\eps_1,\nonumber\\
d^{(\as)}_s&=&-\frac{C\,r\as\eps_1}{9\pi}[3(3-2\ln2) \ln r
+9 ( \ln^2 2 -3 \ln2+3)+ \pi^2]\nonumber\\ &&\approx-\frac{C\,r\as\eps_1(-4.8\ln r-22.5)}{9\pi},\nonumber\\
d^{(\as v^2)}_s&=&\frac{C\,r\as\eps_1}{108m_c^2\pi}[3(27-10 \ln2) \ln r
+(45 \ln^2 2-75 \ln2-79)+5 \pi^2]\nonumber\\ & \approx&-\frac{C\,r\as\eps_1(5.0\ln r-5.0)}{9\pi m_c^2},
\end{eqnarray}
where $C\equiv\frac{(4\pi\alpha)Q_c^2}{2m_c^3}$.
The terms of $\eps_2$ disappear in the expressions because $\eps_2$ is suppressed by a factor of $r$ than $\eps_1$.
The asymptotic behavior of $d^{(\as)}$ is consistent with that in Ref.\cite{Sang:2009jc}.

The asymptotic behavior of the coefficients for $r\rightarrow\infty$ corresponds to the process $\eta_c\rightarrow2\gamma$ are mentioned in Ref.\cite{Sang:2009jc} and given as follows\footnote{Comparing of the previous results of the $\mathcal{O}(v^2)$ corrections with the di-photon decay process for $\eta_c$ and $\chi_c$ \cite{Ma:2002eva,Bodwin:1994jh,Bodwin:2002hg}, we find that the absolute values of our results differ by $1/4$ which originates from the relativistic renormalization expansion (Eq.\ref{eq_nrqcd_amp_ex_asv2}). Therefore the coefficients of the relativistic corrections for the $\eta_c$, $\chi_c$ decay widths shown in Tab.\ref{tab:c_h_decay_2r} differ by $1/2$ from the previous works.}:
\begin{eqnarray}
&&\lim\limits_{r\rightarrow\infty}A^{(0)}=-C,\nonumber\\
&&\lim\limits_{r\rightarrow\infty}A^{(v^2)}=\frac{17}{12m_c^2}C,\nonumber\\
&&\lim\limits_{r\rightarrow\infty}A^{(\as)}=\frac{C\,\as(20-\pi^2)}{6\pi},\nonumber\\
&&\lim\limits_{r\rightarrow\infty}A^{(\as v^2)}=\frac{C\,\as}{216m_c^2\pi}(384\ln2-844+63\pi^2
).
\end{eqnarray}
Note that,
\begin{eqnarray}
\lim\limits_{r\rightarrow\infty}\eps_2=\eps^{\mu\nu\rho\tau}
(\eps^{\ast}_{Q})_{\mu}(\eps^{\ast}_{k})_{\nu}(k^{(2)}_{\rho}p^{(0)}_{\tau}+k^{(0)}_{\rho}p^{(2)}_{\tau})=\eps_1.
\end{eqnarray}
Therefore, the NLO short-distance coefficients in $v^2$ are given by
\begin{eqnarray}
&&\lim\limits_{r\rightarrow\infty}d^{(v^2)}_s=\frac{5C}{12m_c^2}\eps_1
=-\frac{5}{12m_c^2}\lim\limits_{r\rightarrow\infty}d^{(0)},\nonumber\\
&&\lim\limits_{r\rightarrow\infty}d^{(\as v^2)}_s=\frac{C\,\as}{m_c^2\pi}
(\frac{16\ln2}{9}-\frac{31}{54}+\frac{\pi^2}{8}).
\end{eqnarray}
The short-distance in $\mathcal{O}(\as v^2)$ is consistent with Ref.\cite{Jia:2011ah} if we disregard the contribution of the relativistic renormalization in Eq.(\ref{eq_nrqcd_amp_ex_asv2}) which contributes a factor of $1/4$.

\subsection{Matching results for $\chi_{cJ}$}

This subsection presents the matching results for the short-distance coefficients for  $\chi_{cJ}$.


Similar to the $\eta_c$ case, the short-distance coefficients in the orders of $v^2$ and $\as v^2$ for $\chi_{cJ}$ are also written in two parts. All the coefficients are given as follows:
\begin{eqnarray}\label{eq_chicsd}
d^{(0)}_t&=&B^{(0)}\eps_3,\nonumber\\
d^{(v^2)}_t&=&B^{(v^2)}\eps_3+B^{(0)}\eps_4/m_c^2,\nonumber\\
d^{(\as)}_t&=&B^{(\as)}\eps_3,\nonumber\\
d^{(v^2)}_t&=&B^{(\as v^2)}\eps_3+B^{(\as)}\eps_4/m_c^2,
\end{eqnarray}
where
\begin{eqnarray}\label{eq_e3e4def}
&&\eps_3\equiv(\eps^{\ast}_{Q})_{\mu}(\eps^{\ast}_{k})_{\nu}
\mathcal{P}_{\alpha\beta}^{(0)},\nonumber\\
&&\eps_4\equiv(\eps^{\ast}_{Q})_{\mu}(\eps^{\ast}_{k})_{\nu}
\mathcal{P}_{\alpha\beta}^{(2)}.
\end{eqnarray}

The asymptotic behavior  in the limit $r\rightarrow0$ is also considered.
For the $\eps_4$ is  higher order than $\eps_3$ in $r$, then the coefficients are given as follows:
\begin{eqnarray}
\lim\limits_{r\rightarrow0}d^{(0)}_t&=&D\eps_3(g^{\alpha\nu}g^{\beta\mu}-g^{\alpha\mu}g^{\beta\nu}),\\
\lim\limits_{r\rightarrow0}d^{(v^2)}_t&=&-\frac{D\eps_3}{20m_c^2}(11g^{\alpha\nu}g^{\beta\mu}
-11g^{\alpha\mu}g^{\beta\nu}+2g^{\alpha\beta}g^{\mu\nu}),\nonumber\\
\lim\limits_{r\rightarrow0}d^{(\as)}_t&=&\frac{D\as\eps_3}{9\pi}\Big\{(g^{\alpha\nu}g^{\beta\mu}
-g^{\alpha\mu}g^{\beta\nu})[3(3-2\ln2)\ln r + 3(3\ln^2 2 -5 \ln2 +7) \pi^2]\nonumber\\&& + 6g^{\alpha\beta}g^{\mu\nu}(1+2\ln2)\Big\},\nonumber\\
\lim\limits_{r\rightarrow0}d^{(\as v^2)}_t&=&\frac{D\as\eps_3}{540\pi m_c^2}\Big\{ 6g^{\alpha\beta}g^{\mu\nu}[3(1-2\ln2)\ln r+(9\ln^2 2-99\ln2-93)+\pi^2] \nonumber\\
&&+(g^{\alpha\nu}g^{\beta\mu}-g^{\alpha\mu}g^{\beta\nu})[9(45-22\ln2)\ln r +(297\ln^2 2-75\ln2-77+33\pi^2)]\Big\}.\nonumber
\end{eqnarray}

\section{Cross section}\label{sec_cs}

The cross sections of the process $e^+e^-\rightarrow\gamma H$ are relative to the squared amplitudes of the process $\gamma^\ast\rightarrow\gamma H$,
\begin{eqnarray}\label{q_sig}
\sigma(e^+e^-\rightarrow\gamma H)=\frac{1}{2s}\frac{2(D-2)(4\pi \alpha)}{(D-1)s}\int{\Phi_2 \overline{\sum}|\mathcal{M}(\gamma^\ast\rightarrow\gamma H)|^2}.
\end{eqnarray}
$\overline{\sum}$ means obtaining the  sum of NRQCD amplitudes $\mathcal{M}$ over the final-state color and polarization and the average over the ones of the initial states.
Where the differential two-body phase space in D dimensions can be solved:
\begin{eqnarray}\label{q_phs}
\int{\Phi_2}&=&\frac{1}{8\pi}(\frac{4\pi}{s})^\eps(1-\frac{M_{H}^2}{s})^{1-2\eps}
\frac{\Gamma(1-\eps)}{\Gamma(2-2\eps)} \nonumber \\ &\approx&\frac{1}{8\pi}(\frac{4\pi}{s})^\eps(1-r)^{1-2\eps}
\big[1-\frac{r(1-2\eps)}{1-r}\frac{\textbf{q}^2}{m_c^2}\big]\frac{\Gamma(1-\eps)}{\Gamma(2-2\eps)},
\end{eqnarray}
where $M_{H}\approx2E_q$ has been chosen. This expression implies that the two-body phase space contributes another factor of $-r/(1-r)$ to the $v^2$ order cross section. This factor is linearly divergent near the low-energy threshold.

The results of short-distance amplitudes are obtained in  the last section.
Then the cross sections for $\eta_c$ and $\chi_{cJ}$ states  can be obtained  as follows:
\begin{eqnarray}\label{eq_sig_radio}
\sigma=\hat{\sigma}^{(0)}\big[1 + \as c^{10} + (c^{02} + \as c^{12})\langle v^2\rangle\big]\langle0|\mathcal{O}^H|0\rangle,
\end{eqnarray}
where $\hat{\sigma}^{(0)}$ is the LO short-distance cross section, and the matrix element $\langle v^2\rangle$ is defined as follows:
\begin{eqnarray}
\langle v^2\rangle\equiv\frac{\langle0|\mathcal{P}^H|0\rangle}{m_c^2\langle0|\mathcal{O}^H|0\rangle}.
\end{eqnarray}

\subsection{$\eta_c$}

The LO short-distance cross section for $\eta_c$ is given by:
\begin{eqnarray}
\hat{\sigma}^{(0)}_{\eta_c}=\frac{(4\pi\alpha)^3Q_c^4(1-r)}{6\pi m_c s^2}.
\end{eqnarray}

\begin{figure}[ht]
\begin{center}
\includegraphics[width=0.86\textwidth]{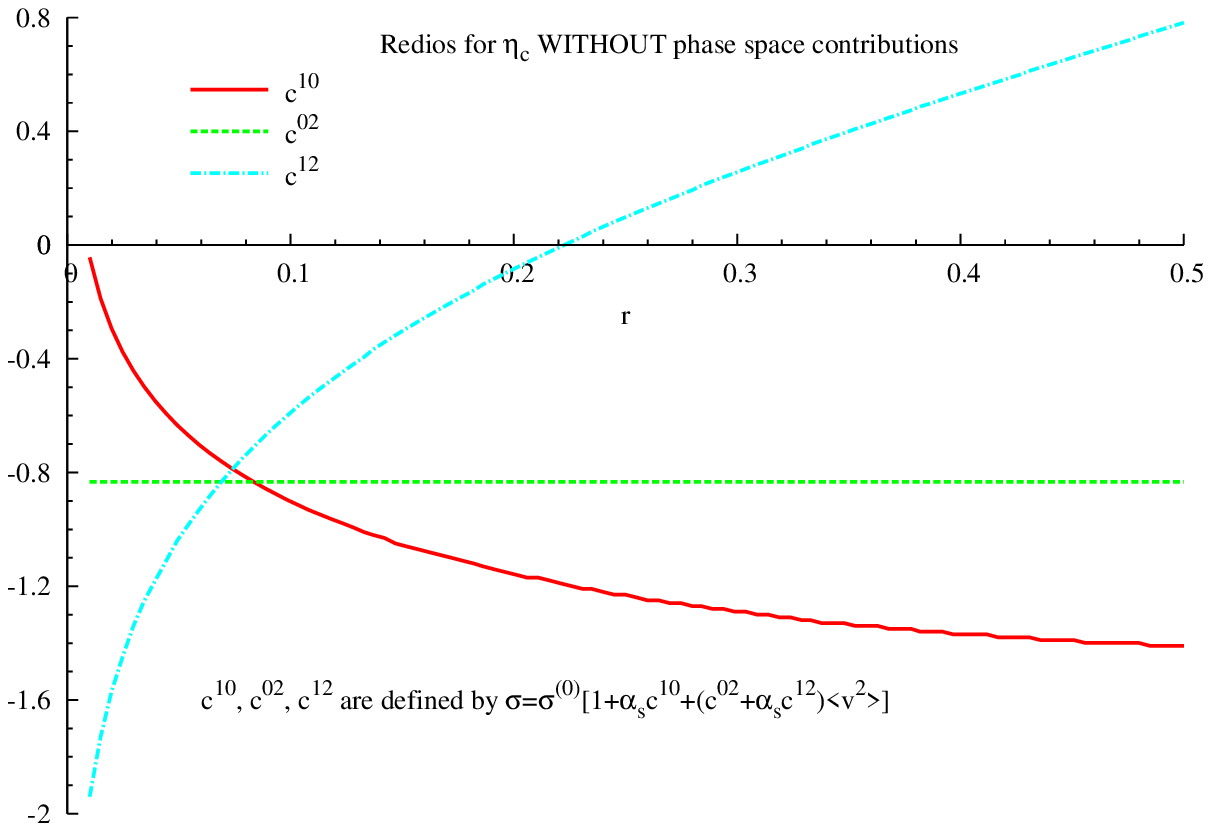}
\includegraphics[width=0.86\textwidth]{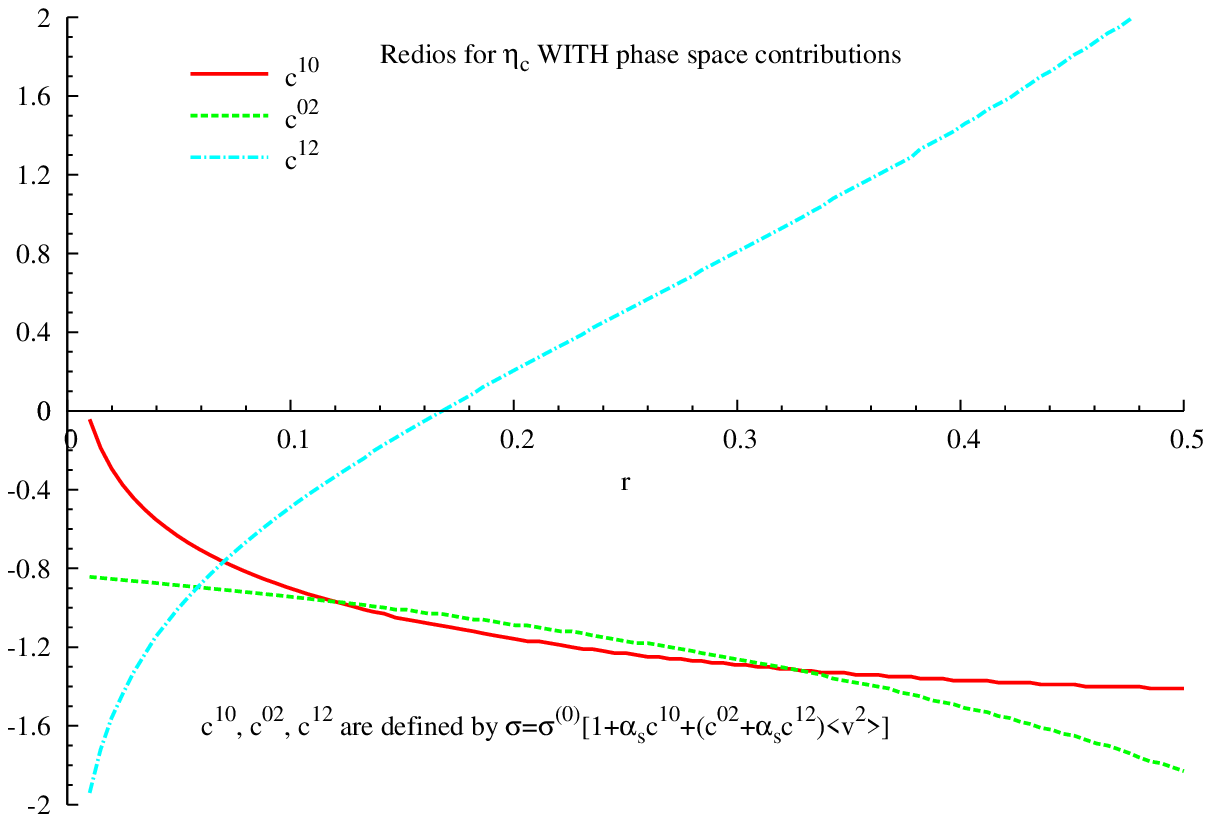}
\end{center}
\caption{\label{fig:etac_c}  The relative ratios for the corrections in the order of $\as$, $v^2$, and $\as v^2$ to the LO cross section for $\eta_c$ production recoiled with a hard photon as a  function of $r$.
The ratios $c^{10}$, $c^{02}$, and $c^{11}$ are defined by the expression $\sigma=\hat{\sigma}^{(0)}\big[1 + \as c^{10} + (c^{02} + \as c^{12})\langle v^2\rangle\big]\langle0|\mathcal{O}^H|0\rangle$.  }
\end{figure}

Fig.\ref{fig:etac_c} shows that the radios $c^{10}$, $c^{02}$, and $c^{11}$ ranging $r$ from $0$ to $0.5$, ie. high-energy to low-energy,  for $\eta_c$ production.
The $\mathcal{O}(\as v^2)$ correction is suppressed by $\as\langle v^2\rangle$ and negligibly contributes to the total cross section at the $r=0.5$.
Tab.\ref{tab:c_near_threshold} presents the asymptotic behaviors of the radios near the threshold. The radio $c^{12}$ for the $\mathcal{O}(\as v^2)$ correction is about $4.8/\pi$ if the corrections from the phase space are not considered, the $\mathcal{O}(\as v^2)$ contribution without the phase space contribution is one fifth of the $\mathcal{O}(\as)$ contribution near the threshold if $\langle v^2\rangle=0.2$.
The phase space brings an additional linear singularity factor that markedly enhances the $\mathcal{O}(v^2)$ and $\mathcal{O}(\as v^2)$ contributions.
However, the total coefficient of the singularity $1/(1-r)$ is  $(4\as/\pi-1)\langle v^2\rangle$ and there is a negative residue singularity.
The $\mathcal{O}(\as v^2)$ corrections becomes significant at the high-energy region and provides  negative contributions under the same sign with the $\mathcal{O}(v^2)$ and $\mathcal{O}(\as)$ ones as Tab.\ref{tab:c_highernergy}. For the B factories  energy  region ( $r\approx0.07$ ), the contributions from $\mathcal{O}(\as v^2)$ corrections are numerically suppressed by $\langle v^2\rangle$ than those from $\mathcal{O}(\as)$ corrections.

Tab.\ref{tab:c_h_decay_2r} lists the corresponding coefficients for the decay process $\eta_c\to\gamma\gamma$. The $\mathcal{O}(\as v^2)$ contribution slightly affects  the decay rate, although our numerically calculated value is slightly larger than that from Ref.\cite{Jia:2011ah}. However, the $\mathcal{O}(\as v^2)$ contribution can re-determine the elements $\langle v^2\rangle$ for the color-singlet $S$-wave states.

\begin{table*}[htbp]\large
\caption{
The asymptotic behaviors of the radios near the threshold. The radios are defined in Eq.(\ref{eq_sig_radio}). 
The last term in each cell of $c^{02}$ and $c^{12}$ originates  from the phase space corrections.
\label{tab:c_near_threshold} }
\centering
\begin{tabular}{c|c|c|c}
\hline
& $\lim\limits_{r\rightarrow1}c^{02}$ & $\lim\limits_{r\to1}c^{10}$  & $\lim\limits_{r\to1}c^{12}$ \\ \hline\hline
$\eta_c$ & $-\frac{5}{6}-\frac{1}{1-r}$ & $-\frac{4}{\pi}$ & $\frac{130}{27\pi}+\frac{4}{\pi(1-r)}$ \\ \hline
$\chi_{c0}$ & $\frac{2}{1-r}-\frac{11}{10}-\frac{1}{1-r}$   & $-\frac{16}{3\pi}$ & $-\frac{32}{3\pi(1-r)}+\frac{160\ln[2(1-r)]-166}{45\pi}+\frac{16}{3\pi(1-r)}$   \\ \hline
$\chi_{c1}$ & $\frac{2}{1-r}-\frac{13}{5}-\frac{1}{1-r}$ & $-\frac{16}{3\pi}$ & $-\frac{32}{3\pi(1-r)}+\frac{160\ln[2(1-r)]+389}{45\pi}+\frac{16}{3\pi(1-r)}$ \\ \hline
$\chi_{c2}$ & $\frac{2}{1-r}-2-\frac{1}{1-r}$ & $-\frac{16}{3\pi}$ & $-\frac{32}{3\pi(1-r)}+\frac{160\ln[2(1-r)]+131}{45\pi}+\frac{16}{3\pi(1-r)}$ \\
\hline
\end{tabular}
\end{table*}

\begin{table*}[htbp]\large
\caption{
The asymptotic behaviors of the radios in the high-energy limit. The radios are defined in Eq.(\ref{eq_sig_radio}). 
The asymptotic results for $c^{10}$ are consistent with Ref.\cite{Sang:2009jc} and for $c^{02}$ are consistent with Ref.\cite{Li:2013nna}.
\label{tab:c_highernergy} }
\centering
\begin{tabular}{c|cl}
\hline
\hline
$\eta_c$ 	&$\lim\limits_{r\rightarrow0}c^{02}$&$=-\frac{5}{6}$	\\ \hline
  &$\lim\limits_{r\rightarrow0}c^{10}$&$=-\frac{2}{9\pi}[3(3-2\ln2)\ln r + 9(\ln^2 2-3\ln2 + 3) +\pi^2]$ \\ &&$\approx-0.34\ln r-1.59$	\\ \hline
  &$\lim\limits_{r\rightarrow0}c^{12}$&$=\frac{1}{27\pi}[3(21-10\ln2)\ln r+15(3\ln^2 2-7\ln2)+28+5\pi^2]$ \\ &&$\approx0.50\ln r+0.31$	\\ \hline\hline
$\chi_{c0}$	&$\lim\limits_{r\rightarrow0}c^{02}_{\chi_{c0}}$&$=-\frac{13}{10}$	\\\hline
  &$\lim\limits_{r\rightarrow0}c^{10}_{\chi_{c0}}$&$=-\frac{2}{9\pi}[3(1-2\ln2)\ln r + 3\ln2(3\ln2 - 11) +\pi^2]$ \\ &&$\approx0.08\ln r+0.61$	\\\hline
  &$\lim\limits_{r\rightarrow0}c^{12}_{\chi_{c0}}$&$=\frac{1}{135\pi}[9(23-26\ln2)\ln r+3\ln2(117\ln2-443)-637+39\pi^2]$ \\ &&$\approx0.11\ln r-2.37$	\\ \hline\hline
$\chi_{c1}$	&$\lim\limits_{r\rightarrow0}c^{02}_{\chi_{c1}}$&$=-\frac{11}{10}$	\\\hline
  &$\lim\limits_{r\rightarrow0}c^{10}_{\chi_{c1}}$&$=-\frac{2}{9\pi}[3(3-2\ln2)\ln r + 3\ln2(3\ln2 - 5)+21 +\pi^2]$ \\ &&$\approx-0.34\ln r-1.75$	\\\hline
  &$\lim\limits_{r\rightarrow0}c^{12}_{\chi_{c1}}$&$=\frac{1}{135\pi}[9(39-22\ln2)\ln r+3\ln2(99\ln2-95)-637+33\pi^2]$ \\ &&$\approx0.50\ln r+1.36$	\\ \hline\hline
$\chi_{c2}$	&$\lim\limits_{r\rightarrow0}c^{02}_{\chi_{c2}}$&$=-\frac{7}{10}$	\\\hline
  &$\lim\limits_{r\rightarrow0}c^{10}_{\chi_{c2}}$&$=-\frac{2}{9\pi}[3(1-2\ln2)\ln r + 3\ln2(3\ln2 + 1)+18 +\pi^2]$ \\ &&$\approx0.08\ln r-2.42$	\\\hline
  &$\lim\limits_{r\rightarrow0}c^{12}_{\chi_{c2}}$&$=\frac{1}{135\pi}[9(17-14\ln2)\ln r+3\ln2(63\ln2+79)+389+21\pi^2]$ \\ &&$\approx0.15\ln r+2.00$	\\\hline
\hline
\end{tabular}
\end{table*}

\begin{table*}[htbp]\large
\caption{ 
The asymptotic behaviors of the radios in the limt of $r\to\infty$. 
The radios are defined in Eq.(\ref{eq_sig_radio}).
These results are corresponding the radios of the two-photon decay rates for $\eta_c$, $\chi_{c0}$ and $\chi_{c2}$. $\chi_{c1}\rightarrow2\gamma$ is forbade therefore the radios are not given for it.
\label{tab:c_h_decay_2r} }
\centering
\begin{tabular}{c|c|c|c}
\hline
& $\lim\limits_{r\rightarrow\infty}c^{02}$ & $\lim\limits_{r\to\infty}c^{10}$  & $\lim\limits_{r\to\infty}c^{12}$ \\ \hline\hline
$\eta_c$ & $-\frac{5}{6}\approx-0.83$ & $\frac{1}{3\pi}[\pi^2-20]\approx-1.1$ & $-\frac{1}{54\pi}[192\ln 2+21\pi^2-212]\approx-0.8$ \\ \hline
$\chi_{c0}$ & $-\frac{11}{6}\approx-1.83$   & $\frac{1}{9\pi}[3\pi^2-28]\approx-0.06$ & $-\frac{1}{90\pi}[320\ln2+65\pi^2-196]\approx-2.36$   \\ \hline
$\chi_{c2}$ & $-\frac{3}{2}=-1.5$ & $-\frac{16}{3\pi}\approx-1.7$ & $-\frac{1}{135\pi}[48\ln2-9\pi^2-1148]\approx2.8$ \\
\hline
\end{tabular}
\end{table*}

\subsection{$\chi_{cJ}$}

The LO short-distance cross section for $\chi_{cJ}$ is calculated as
\begin{eqnarray}\label{eq_losig_chic}
&&\hat{\sigma}^{(0)}_{\chi_{c0}}=\frac{(4\pi\alpha)^3Q_c^4(1-3r)^2}{18\pi m_c^3 s^2(1-r)},\nonumber\\
&&\hat{\sigma}^{(0)}_{\chi_{c1}}=\frac{(4\pi\alpha)^3Q_c^4(1+r)}{3\pi m_c^3 s^2(1-r)},\nonumber\\
&&\hat{\sigma}^{(0)}_{\chi_{c2}}=\frac{(4\pi\alpha)^3Q_c^4(1+3r+6r^2)}{9\pi m_c^3 s^2(1-r)},
\end{eqnarray}

\begin{figure}[ht]
\begin{center}
\includegraphics[width=0.86\textwidth]{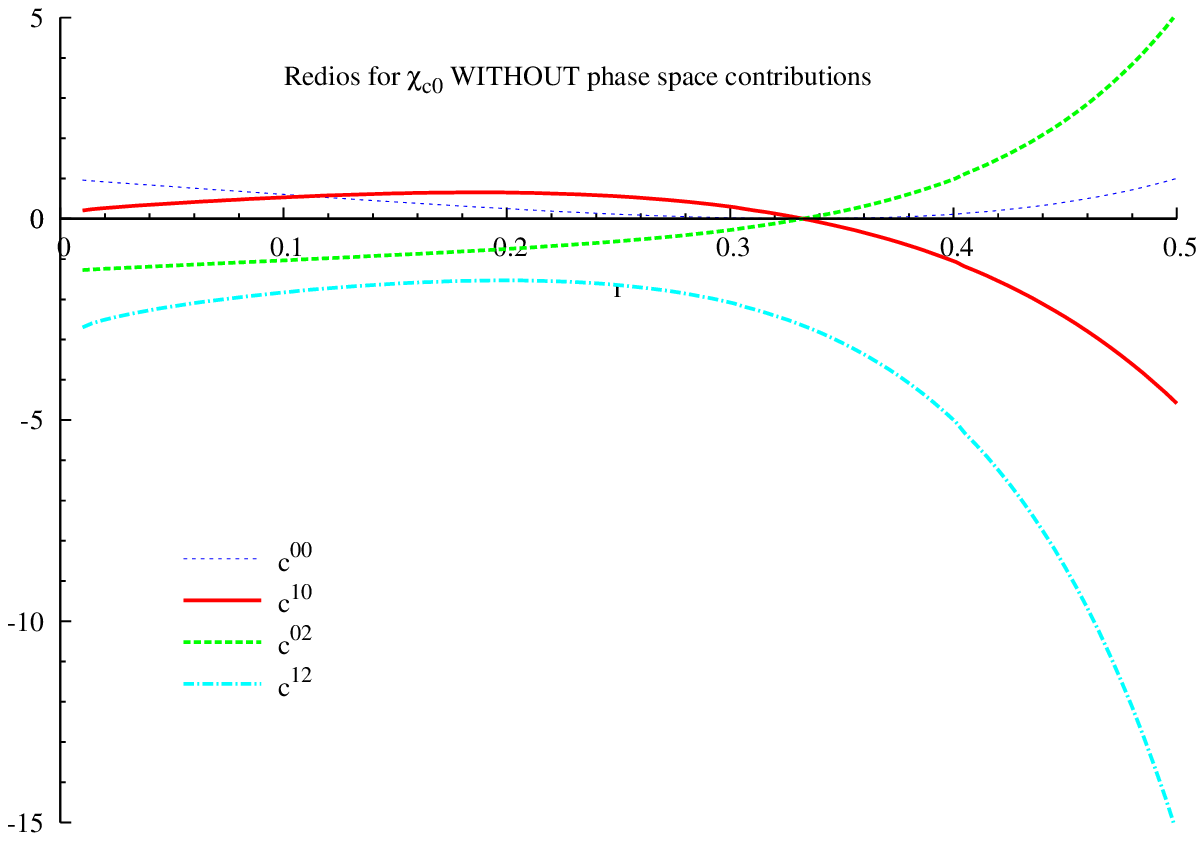}
\includegraphics[width=0.86\textwidth]{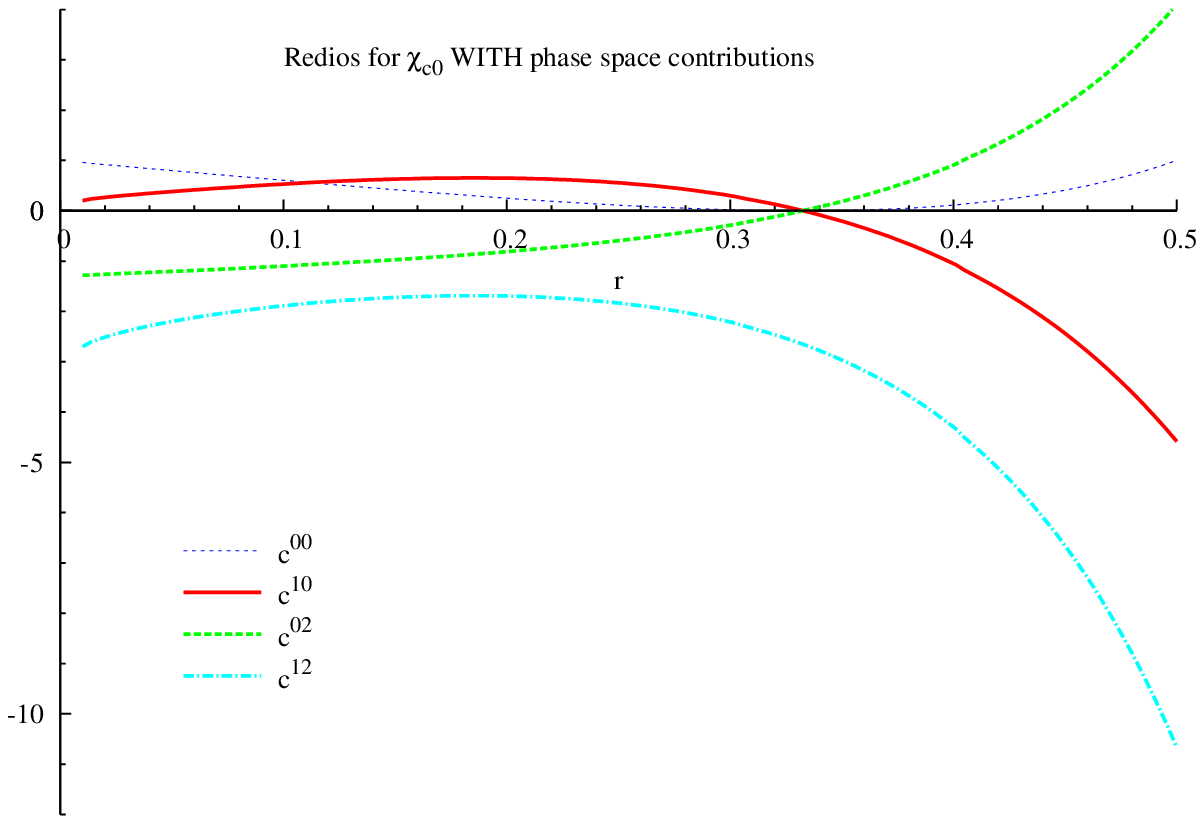}
\end{center}
\caption{\label{fig:chic0_c}  The relative ratios for the corrections in the order of $\as$, $v^2$, and $\as v^2$ to the LO cross section for $\chi_{c0}$ production recoiled with a hard photon as a function of $r$.
The ratios $c^{10}$, $c^{02}$, and $c^{11}$ are defined by the expression $\sigma=\frac{(4\pi\alpha)^3Q_c^4}{18\pi m_c^3 s^2}\big[c^{00} + \as c^{10} + (c^{02} + \as c^{12})\langle v^2\rangle\big]\langle0|\mathcal{O}^H|0\rangle$. }
\end{figure}

For $\chi_{c0}$, the radios may be divergent at $r=1/3$ for the LO short-distance coefficient reach zero at this point as Eq.(\ref{eq_losig_chic}).
Thus, we change Eq.(\ref{eq_sig_radio}) into the following formula to define the radios:
\begin{eqnarray}\label{eq_sig_radio_chic0}
\sigma_{\chi_{c0}}=\frac{(4\pi\alpha)^3Q_c^4}{18\pi m_c^3 s^2}\big[c^{00} + \as c^{10} + (c^{02} + \as c^{12})\langle v^2\rangle\big]\langle0|\mathcal{O}^H|0\rangle.
\end{eqnarray}
The redefined radios are shown as Fig.\ref{fig:chic0_c} and these radios are proportional to the relative short-distance cross sections.
By a rough estimation, the LO cross sections are diluted by the sum of the $\mathcal{O}(\as)$ and $\mathcal{O}(v^2)$ corrections as shown in figure.
Furthermore, the $\mathcal{O}(\as v^2)$ contributes additional negative corrections.
Thus, the total cross sections for the $\chi_{c0}$ process may be small.

\begin{figure}[ht]
\begin{center}
\includegraphics[width=0.86\textwidth]{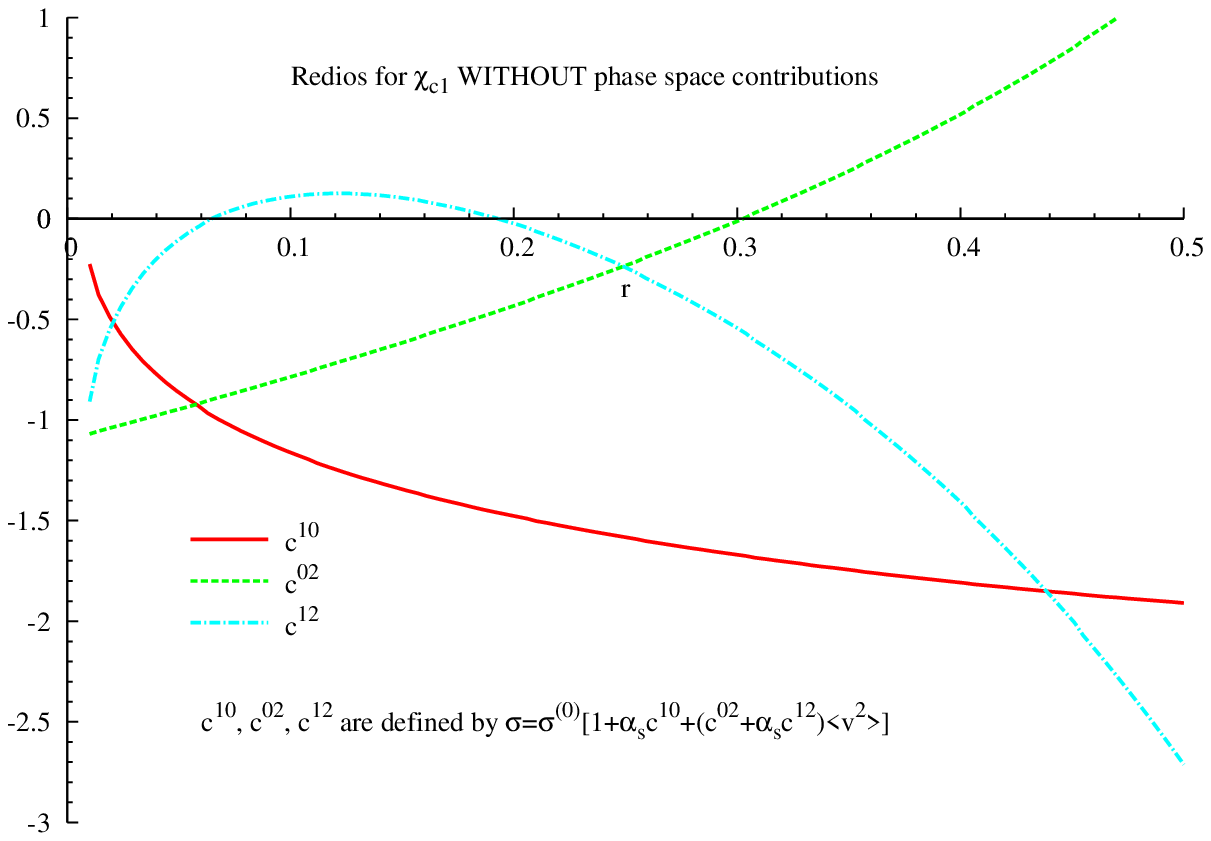}
\includegraphics[width=0.86\textwidth]{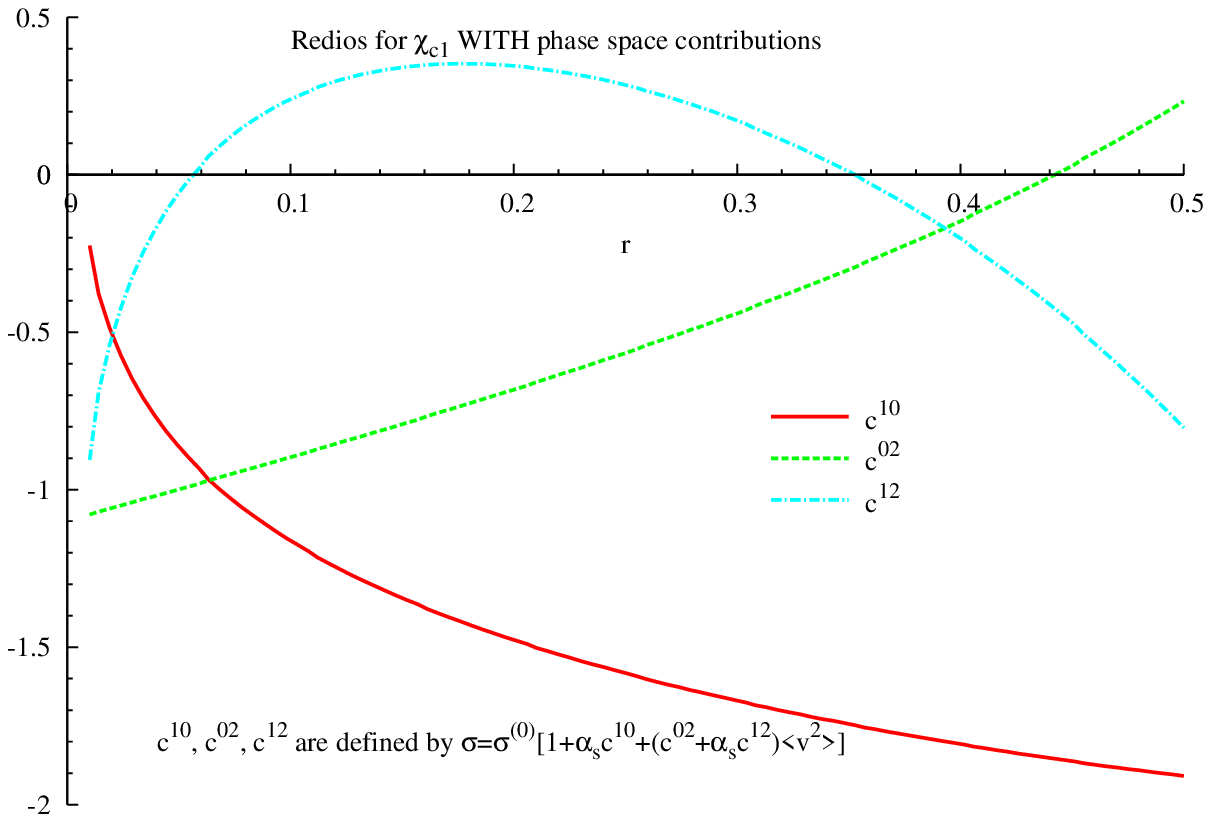}
\end{center}
\caption{\label{fig:chic1_c}  The relative ratios for the corrections in the order of $\as$, $v^2$, and $\as v^2$ to the LO cross section for $\chi_{c1}$ production recoiled with a hard photon as a  function of $r$.
The ratios $c^{10}$, $c^{02}$, and $c^{11}$ are defined by the expression $\sigma=\hat{\sigma}^{(0)}\big[1 + \as c^{10} + (c^{02} + \as c^{12})\langle v^2\rangle\big]\langle0|\mathcal{O}^H|0\rangle$.  }
\end{figure}

\begin{figure}[ht]
\begin{center}
\includegraphics[width=0.86\textwidth]{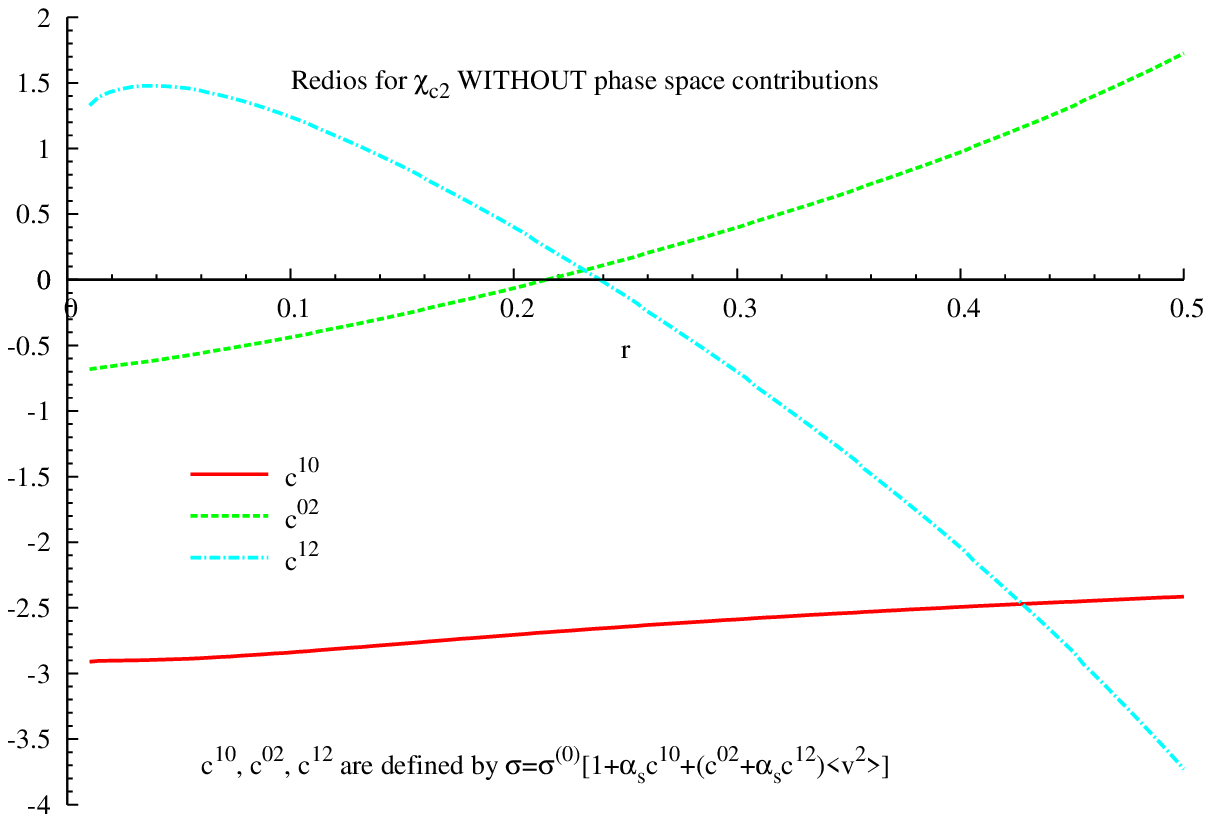}
\includegraphics[width=0.86\textwidth]{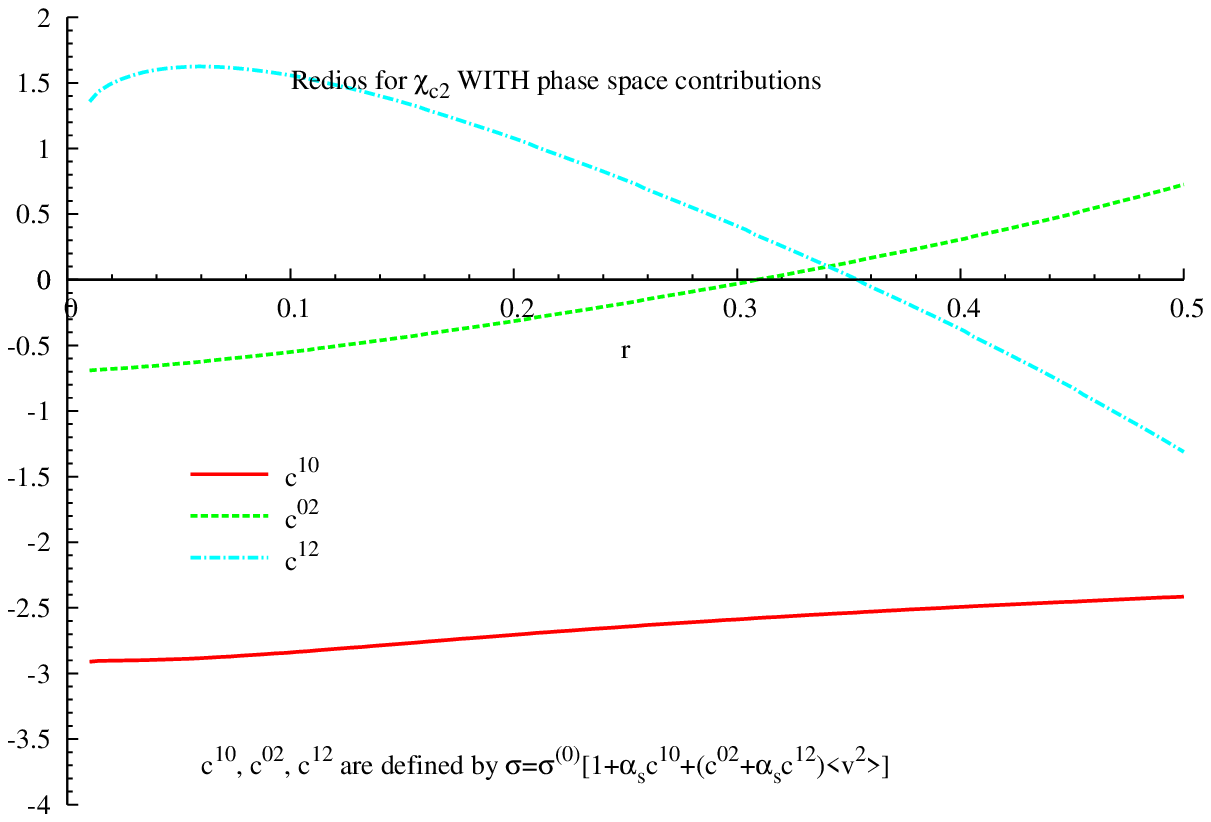}
\end{center}
\caption{\label{fig:chic2_c}  The relative ratios for the corrections in the order of $\as$, $v^2$, and $\as v^2$ to the LO cross section for $\chi_{c2}$ production recoiled with a hard photon as a function of $r$.
The ratios $c^{10}$, $c^{02}$, and $c^{11}$ are defined by the expression $\sigma=\hat{\sigma}^{(0)}\big[1 + \as c^{10} + (c^{02} + \as c^{12})\langle v^2\rangle\big]\langle0|\mathcal{O}^H|0\rangle$.  }
\end{figure}

The radios for $\chi_{c1}$ and $\chi_{c2}$ processes are shown in  Fig.\ref{fig:chic1_c} and Fig.\ref{fig:chic2_c}, respectively.
In the low-energy region ($0.3<r<0.5$), the $\mathcal{O}(\as v^2)$ corrections contribute the most.
Meanwhile the $\mathcal{O}(v^2)$ and $\mathcal{O}(\as v^2)$ corrections increase as  fast as  $r$, and they have different signs.
As shown in Tab.\ref{tab:c_near_threshold}, the behaviors of the radios are similar for all the $P$-wave states.
The radio $c^{02}$ for the $\mathcal{O}(v^2)$ correction near the threshold has an additional linear singularity $1/(1-r)$ for the LO cross section.
The radio $c^{10}$ for the $\mathcal{O}(\as)$ corrections is a negative constant. In other words, the $\mathcal{O}(\as)$ contribution has the same rate as the corresponding LO cross section for different $\chi_{cJ}$ states.
For the $\mathcal{O}(\as v^2)$ corrections, a logarithmic singularity term $\ln(1-r)$ aside from the linear singularity term also exists.
We sum the linear singularity in the $\mathcal{O}(v^2)$ corrections and $\mathcal{O}(\as v^2)$ corrections and obtain the coefficient of the linear singularity as $(1-16\as/3/\pi)\langle v^2\rangle\approx-0.5\langle v^2\rangle$. The coefficient of the residual linear singularity is similar to that of $\eta_c$. The linear singularity half originates from the phase space.
For the high-energy region,  we consider the suppressed factor $\as\langle v^2\rangle$ to the $\mathcal{O}(v^2)$ corrections.
The numerical results in the high-energy approximation in Tab.\ref{tab:c_highernergy} show that the corrections are much smaller than the $\as$ and $v^2$ corrections.

The radios corresponding to the two-photon decay for $\chi_{c0}$ and $\chi_{c2}$ are also given in Tab.\ref{tab:c_h_decay_2r}.
By the rough estimation, we select $\as$ and $\langle v^2\rangle$ as a range of $0.2\sim0.3$. Therefore the $\mathcal{O}(\as v^2)$ corrections contributes $10\%\sim20\%$ to the LO decay rate for $\chi_{c0}\rightarrow2\gamma$ or $\chi_{c2}\rightarrow2\gamma$.
These $\mathcal{O}(\as v^2)$ corrections may also significantly affect fitting to the element $\langle v^2\rangle$ for $\chi_{cJ}$.

\section{Numerical results and discussion}\label{sec:numer}

In this section, we revisit the numerical calculations to the cross sections.
In our numerical calculation, the total cross sections strongly depend on the input parameters (e.g.,  mass of the charm quark, long distance matrix elements, and the strong-coupling constant). The relativistic matrix elements can hardly be determined.
In the consequent calculations for the $\eta_c(1S)$, $\chi_{cJ}(1P)$ process, we select  the fine structure constant $\alpha=1/137$ and the charm quark mass as
\begin{eqnarray}
m_c=1.5\pm0.1GeV,
\end{eqnarray}
for both the $\eta_c(1S)$ and $\chi_{cJ}(1P)$ process. The strong-coupling constant is chosen as
\begin{eqnarray}
\as=0.23\pm0.03.
\end{eqnarray}
The matrix elements $\langle v^2\rangle$ are chosen as
\footnote{For the $P$-wave states, we can use our new up to $\mathcal{O}(\as v^2)$ results for $\chi_{c0}\to\gamma\gamma$ and $\chi_{c2}\to\gamma\gamma$ to fit $\langle v^2\rangle$:
\begin{eqnarray}
&&\Gamma[\chi_{c0}\to\gamma\gamma]=\frac{6\pi Q_c^4 \alpha^2}{m_c^4}\langle\mathcal{O}^{\chi_{c0}}\rangle[1-0.06\as-(1.83+2.36\as)\langle v^2\rangle^{\chi_{c0}}-3a_8],\nonumber\\
&&\Gamma[\chi_{c2}\to\gamma\gamma]=\frac{8\pi Q_c^4 \alpha^2}{5m_c^4}\langle\mathcal{O}^{\chi_{c2}}\rangle[1-1.7\as-(1.5-2.8\as)\langle v^2\rangle^{\chi_{c2}}-2.3a_8-1.7a_F]. \nonumber
\end{eqnarray}
The color-octet contributions in the above formula originate from  Ref.\cite{Ma:2002eva}.
In the estimation, we ignore the $v^2$ corrections to the elements and assume $\langle\mathcal{O}^{\chi_{c2}}\rangle=5\langle\mathcal{O}^{\chi_{c0}}\rangle$ and $\langle v^2\rangle^{\chi_{c}}\equiv\langle v^2\rangle^{\chi_{c0}}=\langle v^2\rangle^{\chi_{c2}}$.
If we take $a_8=a_F=0.1$ then $\langle v^2\rangle^{\chi_c}=0.32\pm0.04$ is obtained. If we
ignore the contributions of the $a_8$ and $a_F$ terms,
then $\langle v^2\rangle^{\chi_c}=0.21\pm0.03$.
Therefore $\langle v^2\rangle^{\chi_c}=0.2\pm0.1$ are compatible for these results.
In this study, we select $\as=0.23\pm0.03$ and the di-photon decay width for $\chi_{c0}$ and $\chi_{c2}$ are $(2.23\pm0.17)\times10^{-4}{\rm MeV/c}$ and $(2.59\pm0.16)\times10^{-4}{\rm MeV/c}$, respectively, cited from PDG\cite{Beringer:1900zz}.}
\begin{eqnarray}
&&\langle v^2\rangle^{\eta_c}=0.15\pm0.1,\nonumber\\
&&\langle v^2\rangle^{\chi_{cJ}}=0.20\pm0.1,
\end{eqnarray}
The LO long-distance matrix elements are obtained from the radial wave functions at the origin in the potential model calculations \cite{Eichten:1995ch} with the replacements
\begin{eqnarray}
\langle0|\mathcal{O}^{\eta_c(nS)}|0\rangle&=&\frac{2N_c|R_{nS}(0)|^2}{4\pi},\nonumber\\
\langle0|\mathcal{O}^{\chi_{c0}(mP)}|0\rangle&=&\frac{6N_c|R_{mP}^\prime(0)|^2}{4\pi},
\end{eqnarray}
and
\begin{eqnarray}
\langle0|\mathcal{O}^{\chi_{cJ}(mP)}|0\rangle&=&(2J+1)(1+\mathcal{O}(v^2))\langle0|\mathcal{O}^{\chi_{c0}(mP)}|0\rangle\nonumber\\
&\approx&(2J+1)\langle0|\mathcal{O}^{\chi_{c0}(mP)}|0\rangle.
\end{eqnarray}
In the last step, we ignore the $\mathcal{O}(v^2)$ term to simplify the input parameters.
The results markedly depend on the selections of the wave functions at the origin.
Studies in Ref.\cite{Chao:2013cca,Li:2009ki} have adopted two sets of wave functions  at the origin  with large gaps.
We re-estimate the wave functions  at the origin  by averaging the two sets of wave  functions with the uncertainties in Tab.\ref{tab:wf}.
\begin{table*}[htbp]
\caption{ The wave functions  at the origin \cite{Eichten:1995ch}. The two sets represent the calculations from the Cornell potential and the B-T potential.
"Re-est" are averaged from the two sets of functions with the uncertainties.
\label{tab:wf} }
\centering
\begin{tabular}{c|ccccc}
\hline
& $1S({\rm GeV}^3)$ & $2S({\rm GeV}^3)$  & $3S({\rm GeV}^3)$  & $1P({\rm GeV}^5)$  & $2P({\rm GeV}^5)$ \\ \hline
Cornell & 1.454   & 0.927 & 0.791 & 0.131 & 0.186  \\
B-T & 0.81   &  0.529 & 0.455 & 0.075 & 0.102 \\
Re-est & $1.132\pm0.322$   &  $0.728\pm0.199$ & $0.623\pm0.168$ & $0.103\pm0.028$ & $0.144\pm0.042$ \\
\hline
\end{tabular}
\end{table*}
The wave functions at the origin for $4S$ and $3P$ states are estimated like Ref.\cite{Li:2013nna} as
\begin{eqnarray}
&&R_{4S}=2R_{3S}-R_{2S}=0.518\pm0.391~{\rm GeV}^3,\nonumber\\
&&R^\prime_{3P}=(R^\prime_{1P}+R^\prime_{2P})/2=0.124\pm0.025~{\rm GeV}^5.
\end{eqnarray}

In the BESIII  energy  region, the corrections from the phase space are  significant for the cross sections.
Hovever, the contributions cannot be determined because of the non-perturbative effects.
In the previous works, two different strategies are used to  to remedy the non-perturbative effects from the phase space integrand, a extra factor is introduced in Ref.\cite{Chao:2013cca} and the charm quark mass is set to half of the meson mass in Ref.\cite{Li:2013nna}.
Furthermore, as stated in Ref.\cite{Beneke:1997qw,Beneke:1999gq,Jia:2009np}, the $v^2$ corrections from the phase space, which are related to the terms in the short-distance cross section expansion different with that in the sub-amplitude expansion, could be resummed to all orders in $v^2$ by the 'shape functions' method.
In this paper, we calculate the contributions from the phase space just by a simplified expansion by Eq.(\ref{q_phs}).
Therefore, we analyze the cross sections with and without the phase space corrections for  comparative and referential purposes.

\begin{figure}[ht]
\begin{center}
\includegraphics[width=0.86\textwidth]{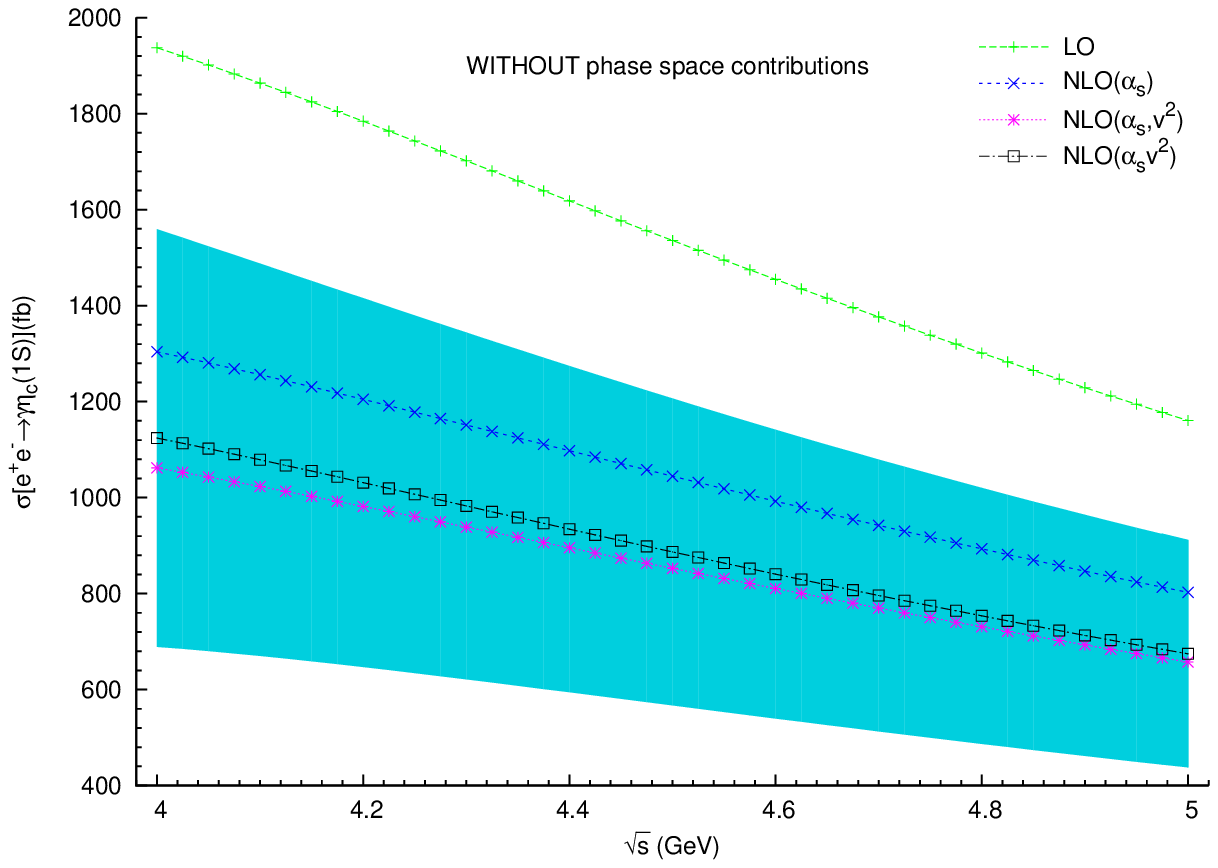}
\includegraphics[width=0.86\textwidth]{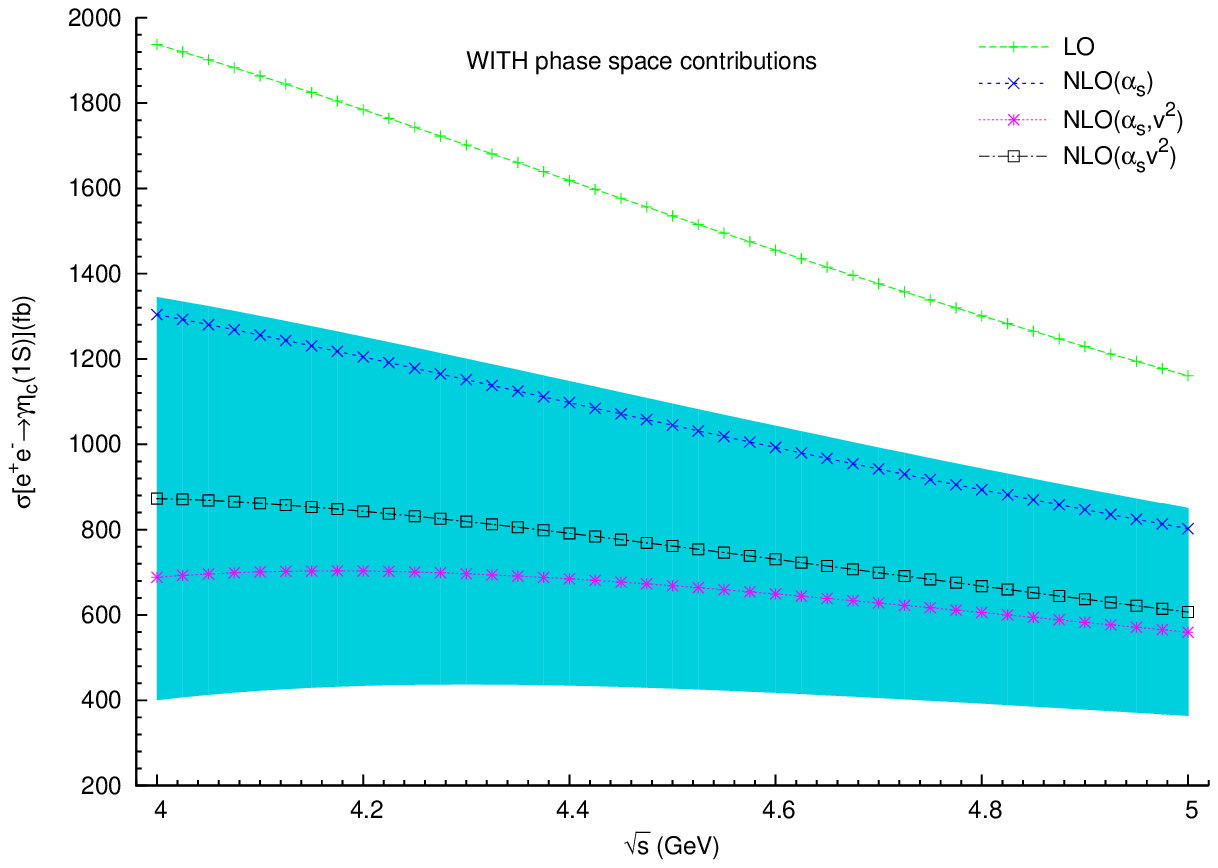}
\end{center}
\caption{\label{fig:etac_cs}  The cross sections of the $\eta_c(1S)$ process at the BESIII  energy region.
The uncertainties for the total cross sections come from the uncertainties of $m_c$, $\as$, $\langle v^2\rangle$, and the wave functions at the origin. }
\end{figure}

Tab.\ref{tab:etaBESB} presents the total cross sections of up to $\as v^2$ order and the corresponding uncertainties for the $\eta_c$ process. And Fig.\ref{fig:etac_cs} presents the corresponding  cross sections at the BESIII  energy region.
The uncertainties for the total cross sections come from the uncertainties of $m_c$, $\as$, $\langle v^2\rangle$, and the wave functions at the origin.
The phase space  reduces the numerical results by a factor of $25\%\sim10\%$ and  enhances the uncertainties by a factor of $35\%\sim25\%$ in the BESIII  energy  region $4\sim5{\rm GeV}$.
The $\mathcal{O}(\as v^2)$ corrections negligibly contribute to the $\eta_c$ process.
Numerical simulations reveal that these corrections are approximately one-eighth and one-tenth of the $\mathcal{O}(\as)$ contributions in the energy regions of the B- factories and BESIII, respectively.
\begin{table*}[htbp]\scriptsize
\caption{The total cross sections in
~$fb$~ up to $\as v^2$ order of $e^+e^- \to \eta_c (nS) + \gamma$
with $n=1,2,3,4$ in the BESIII and B-factories  energy  region.
¡°WP¡± and ¡°OP¡± indicate considering or ignoring the phase space contributions, respectively.
The uncertainties in each cell originate from the uncertainties
of the wave functions at the origin, $\as$, $\langle v^2\rangle$, and charm quark mass
$m_c$ in turns.
For the excited states, we select  the charm quark mass as the half of the meson mass
in the calculations, therefore there are no $m_c$ uncertainties.
The masses of $\eta_c(nS)$ are selected as $3.639{\rm GeV}$, $3.994{\rm GeV}$, and
 $4.250{\rm GeV}$ for $n=2,3,4$ respectively\cite{Li:2009zu,Beringer:1900zz}.
\label{tab:etaBESB} }

\centering
\begin{tabular}{c|ccc}

\hline
$\sqrt s$(GeV)&  4.25  & 4.50  & 4.75  \\ \hline
$1S$ OP & $1007\pm286\pm68\pm114\pm198$ & $887\pm252\pm60\pm105\pm155$ & $775\pm220\pm52\pm95\pm123$  \\
$1S$ WP & $832\pm237\pm57\pm231\pm209$  & $762\pm217\pm52\pm189\pm162$& $683\pm194\pm46\pm156\pm129$  \\
\hline
$2S$ OP &  $282\pm77\pm17\pm26$ & $284\pm77\pm18\pm29$ & $269\pm74\pm18\pm30$  \\
$2S$ WP &  $99\pm27\pm5\pm117$  & $155\pm42\pm10\pm93$& $176\pm48\pm12\pm76$  \\
\hline
$3S$ OP    & $101\pm27\pm5\pm7$ & $142\pm38\pm8\pm12$ & $153\pm41\pm9\pm14$  \\
$3S$ WP     &$-74\pm20\pm5\pm95$ & $19\pm5\pm0.4\pm73$& $65\pm18\pm4\pm58$  \\
\hline
$4S$ OP     &  & $58\pm44\pm3\pm4$ & $81\pm64\pm5\pm7$  \\
$4S$ WP    &   & $-52\pm39\pm4\pm59$& $6\pm4\pm0.1\pm46$  \\
\hline\hline
$\sqrt s$(GeV) & 5.00  & 10.6 & 11.2  \\ \hline
$1S$ OP  & $674\pm192\pm45\pm85\pm100$ & $55\pm16\pm2\pm8\pm5$ & $45\pm13\pm2\pm7\pm4$ \\
$1S$ WP  & $607\pm173\pm41\pm130\pm103$ & $54\pm15\pm2\pm9\pm5$ & $44\pm13\pm2\pm7\pm4$ \\
\hline
$2S$ OP  & $247\pm68\pm17\pm29$ & $25\pm7\pm1\pm4$ & $20\pm6\pm1\pm3$ \\
$2S$ WP  & $179\pm49\pm13\pm63$ & $24\pm7\pm1\pm4$ & $20\pm5\pm1\pm4$ \\
\hline
$3S$ OP   & $151\pm41\pm10\pm16$ & $18\pm5\pm1\pm3$ & $15\pm4\pm1\pm2$ \\
$3S$ WP   & $87\pm23\pm6\pm48$ & $18\pm5\pm1\pm3$ & $15\pm4\pm1\pm3$ \\
\hline
$4S$ OP   & $93\pm70\pm6\pm9$ & $14\pm11\pm1\pm2$ & $12\pm9\pm1\pm2$ \\
$4S$ WP   & $36\pm27\pm2\pm37$ & $13\pm10\pm1\pm3$ & $11\pm8\pm1\pm2$ \\
\hline
\end{tabular}
%
\end{table*}




Tab.\ref{tab:chic0BESB} presents the total cross sections up to $\as v^2$ order for $\chi_{c0}$ process with the uncertainties. The positive $\mathcal{O}(\as)$ corrections and negative $\mathcal{O}(v^2)$ corrections cancel to each other in the BESIII  energy  region\cite{Li:2013nna}, but $\mathcal{O}(\as)$ parts also contribute negative corrections which decrease the LO cross sections significantly even to a negative values.
And the uncertainties are too large compared with the central values to give a reliable predictions for the $\chi_{c0}$ processes in the BESIII  energy  region.

\begin{table*}[htbp]\small
\caption{
The total cross sections in ~$fb$~ up to $\as v^2$ order of $e^+e^- \to \chi_{c0}(nP) + \gamma$ with $n=1,2,3$ in the BESIII and B-factories energy region. ¡°WP¡± and ¡°OP¡± indicate considering or ignoring the phase space contributions, respectively. 
The uncertainties in each cell come from the uncertainties of the wave functions at the origin, $\as$, $\langle v^2\rangle$, and charm quark mass $m_c$ in turns. For the excited states, we select  the charm quark mass as the half of the meson mass in the calculations, therefore there are no $m_c$ uncertainties.
The mass of $\chi_{c0}(nP)$ is selected as $3.918{\rm GeV}$ and $4.131{\rm GeV}$ for $n=2,3$ respectively\cite{Li:2009zu,Beringer:1900zz}.
\label{tab:chic0BESB}}
\centering
\begin{tabular}{c|ccc}
\hline
$\sqrt s$(GeV)  & 4.25  & 4.50  & 4.75  \\ \hline
$1P$ OP &  $17.4\hspace{-0.06cm}\pm\hspace{-0.06cm}4.7\hspace{-0.06cm}\pm\hspace{-0.06cm}15.8\hspace{-0.06cm}\pm\hspace{-0.06cm}10.8\hspace{-0.06cm}\pm\hspace{-0.06cm}39.2$ & $-5.1\hspace{-0.06cm}\pm\hspace{-0.06cm}1.4\hspace{-0.06cm}\pm\hspace{-0.06cm}7.2\hspace{-0.06cm}\pm\hspace{-0.06cm}2.2\hspace{-0.06cm}\pm\hspace{-0.06cm}13.9$ & $-8.8\hspace{-0.06cm}\pm\hspace{-0.06cm}2.4\hspace{-0.06cm}\pm\hspace{-0.06cm}3.3\hspace{-0.06cm}\pm\hspace{-0.06cm}1.0\hspace{-0.06cm}\pm\hspace{-0.06cm}4.1$  \\
$1P$ WP &  $18.3\hspace{-0.06cm}\pm\hspace{-0.06cm}5.0\hspace{-0.06cm}\pm\hspace{-0.06cm}13.9\hspace{-0.06cm}\pm\hspace{-0.06cm}11.3\hspace{-0.06cm}\pm\hspace{-0.06cm}35.8$& $-3.6\hspace{-0.06cm}\pm\hspace{-0.06cm}1.0\hspace{-0.06cm}\pm\hspace{-0.06cm}6.5\hspace{-0.06cm}\pm\hspace{-0.06cm}3.0\hspace{-0.06cm}\pm\hspace{-0.06cm}13.6$& $-8.0\hspace{-0.06cm}\pm\hspace{-0.06cm}2.2\hspace{-0.06cm}\pm\hspace{-0.06cm}3.0\hspace{-0.06cm}\pm\hspace{-0.06cm}0.5\hspace{-0.06cm}\pm\hspace{-0.06cm}4.5$ \\
\hline
$2P$ OP     & $2558\pm746\pm308\pm933$ & $552\pm161\pm83\pm177$ & $170\pm49\pm32\pm51$    \\
$2P$ WP     & $1174\pm723\pm202\pm540$ & $428\pm174\pm60\pm114$& $141\pm58\pm25\pm37$ \\
\hline
$3P$ OP     &  & $1331\pm268\pm163\pm479$ & $320\pm64\pm48\pm102$  \\
$3P$ WP     &  & $932\pm188\pm108\pm280$& $248\pm50\pm35\pm66$ \\
\hline\hline
$\sqrt s$(GeV)  & 5.00  & 10.6 &11.2 \\ \hline
$1P$ OP  & $-7.3\pm2.0\pm1.4\pm2.2\pm0.$ & $1.6\pm0.4\pm0.\pm0.5\pm0.4$& $1.4\pm0.4\pm0.\pm0.4\pm0.3$ \\
$1P$ WP  & $-6.9\pm1.9\pm1.4\pm2.0\pm0.5$ & $1.6\pm0.4\pm0.\pm0.5\pm0.4$& $1.3\pm0.4\pm0.\pm0.4\pm0.3$ \\
\hline
$2P$ OP     & $58\pm17\pm15\pm18$ & $0.7\pm0.2\pm0.\pm0.3$ & $0.6\pm0.2\pm0.\pm0.2$   \\
$2P$ WP     & $51\pm21\pm12\pm15$ & $0.6\pm0.3\pm0.\pm0.3$  & $0.6\pm0.2\pm0.\pm0.2$\\
\hline
$3P$ OP     & $104\pm21\pm20\pm32$ & $0.4\pm0.1\pm0.\pm0.2$ & $0.4\pm0.1\pm0.\pm0.2$ \\
$3P$ WP     & $87\pm17\pm15\pm23$ & $0.4\pm0.1\pm0.\pm0.2$ & $0.4\pm0.1\pm0.\pm0.2$\\
\hline
\end{tabular}
\end{table*}

\begin{figure}[ht]
\begin{center}
\includegraphics[width=0.86\textwidth]{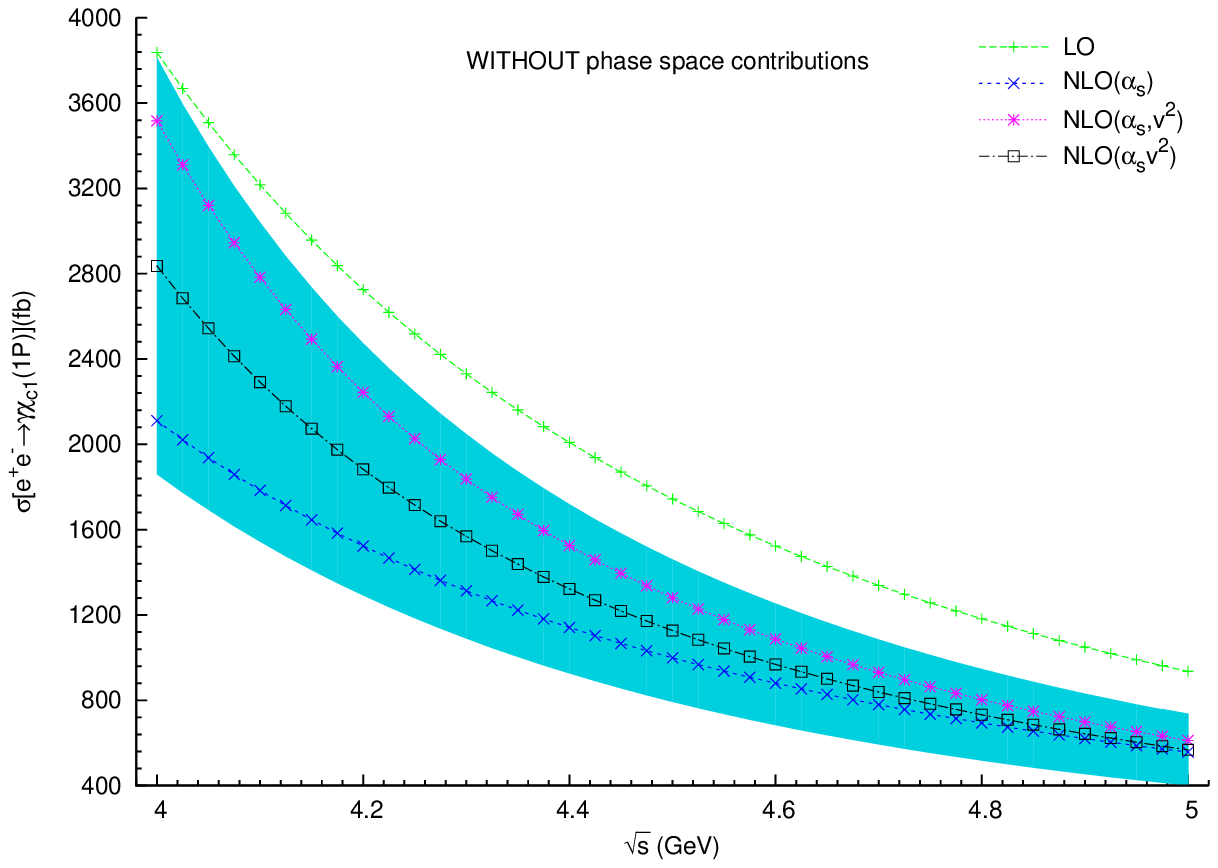}
\includegraphics[width=0.86\textwidth]{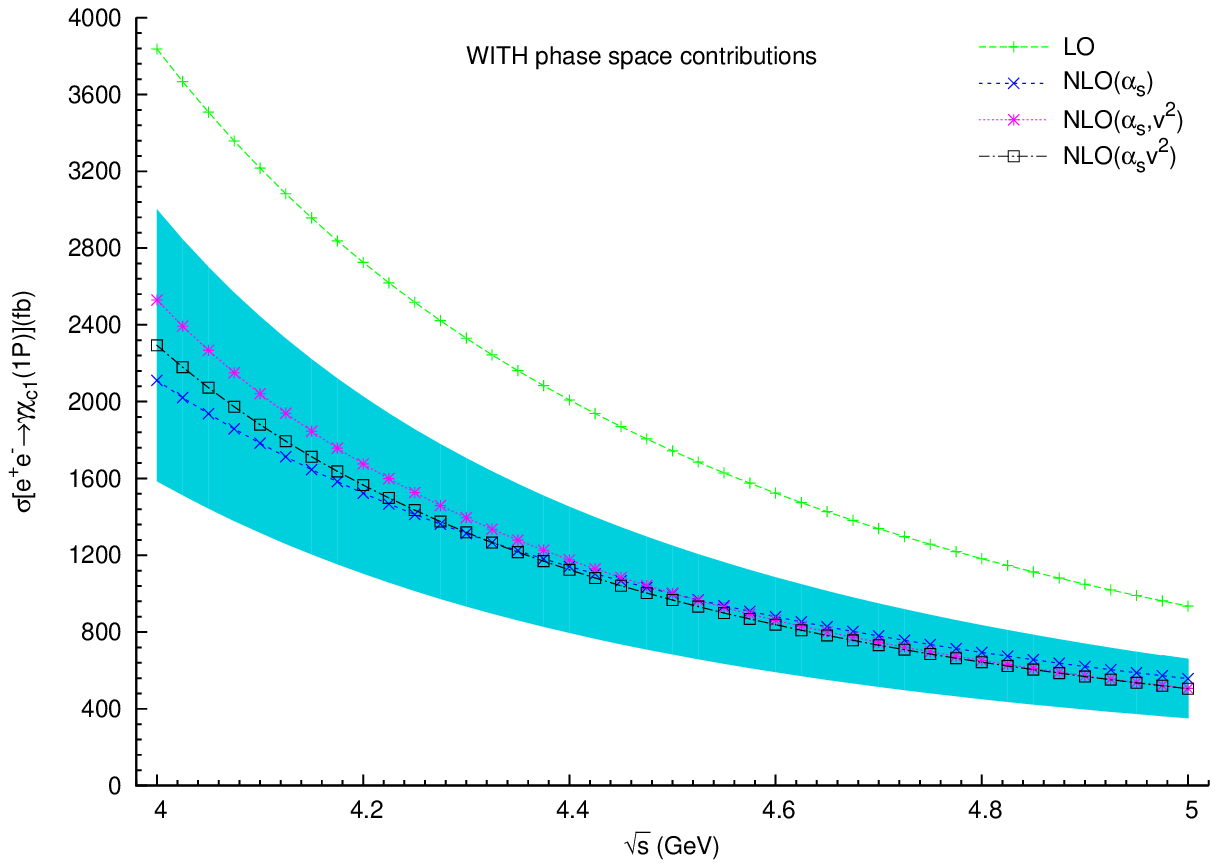}
\end{center}
\caption{\label{fig:chic1_cs}  The cross sections of the $\chi_{c1}(1P)$ process at the BESIII  energy  region.
The uncertainties for the total cross sections come from the uncertainties of $m_c$, $\as$, $\langle v^2\rangle$, and the wave functions at the origin. }
\end{figure}
\begin{figure}[ht]
\begin{center}
\includegraphics[width=0.86\textwidth]{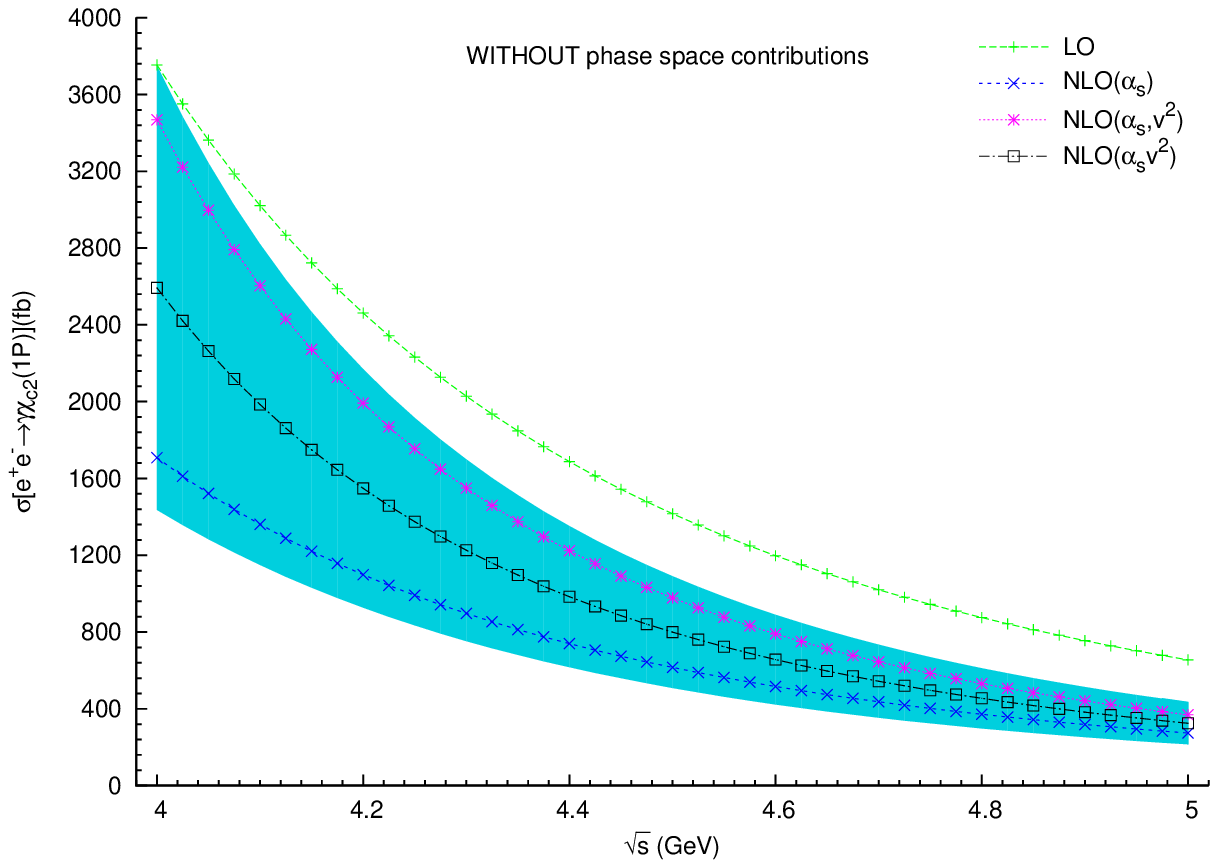}
\includegraphics[width=0.86\textwidth]{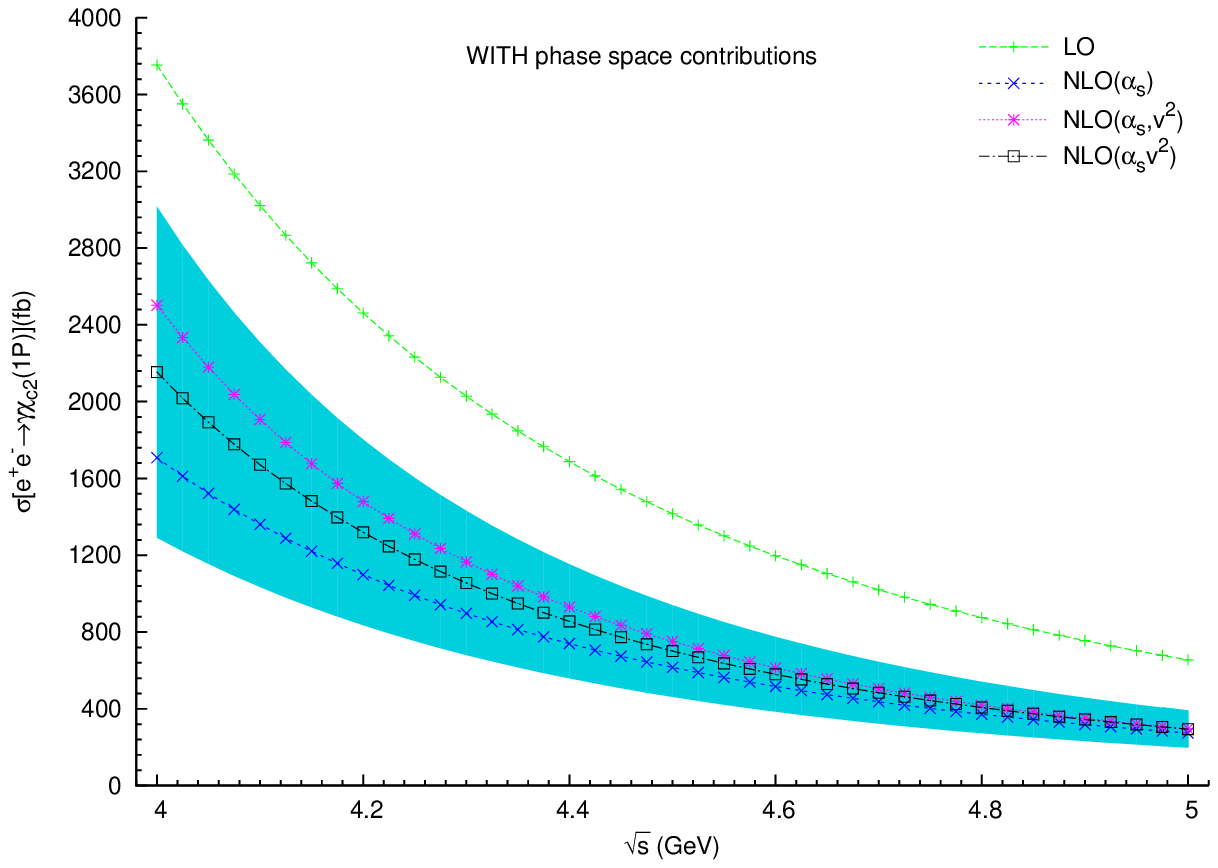}
\end{center}
\caption{\label{fig:chic2_cs}  The cross sections of the $\chi_{c2}(1P)$ process at the BESIII energy region.
The uncertainties for the total cross sections come from the uncertainties of $m_c$, $\as$, $\langle v^2\rangle$, and the wave functions at the origin. }
\end{figure}

Tab.\ref{tab:chic1BESB} and Tab.\ref{tab:chic2BESB} present the total cross sections up to the $\as v^2$ order for the  $\chi_{c1}$ and $\chi_{c2}$ processes, respectively, with the uncertainties. In the BESIII  energy  region, they exhibit similar tends. In addition,  Fig.\ref{fig:chic1_cs} and Fig.\ref{fig:chic2_cs} show that the total cross sections for $\chi_{c2}$ decrease slightly faster than those for $\chi_{c1}$ as the energy increase. The $\mathcal{O}(\as v^2)$ contributions are in behalf of the $\mathcal{O}(v^2)$ ones in this region as discussed in Sec.\ref{sec_cs}.
The phase space corrections  reduce the total cross sections by a factor of $10\%\sim20\%$ in the BESIII  energy  region for both the  $\chi_{c0}$ and $\chi_{c1}$ processes. The corresponding uncertainties  markedly decrease.
From the tables, the $\chi_{c1}$ and $\chi_{c2}$ states will be found in the BESIII  energy  region even if  the lower bound of the numerical values is adopted for  the cross sections.

\begin{table*}[htbp]\small
\caption{
The total cross sections in ~$fb$~ up to $\as v^2$ order of $e^+e^- \to \chi_{c1}(nP) + \gamma$ with $n=1,2,3$ in the BESIII and B-factories energy region. ¡°WP¡± and ¡°OP¡± indicate considering or ignoring the phase space contributions, respectively.
The uncertainties in each cell come from the uncertainties of the wave functions at the origin, $\as$, $\langle v^2\rangle$, and charm quark mass $m_c$ in turns. For the excited states, we select  the charm quark mass as the half of the meson mass in the calculations, therefore there are no $m_c$ uncertainties.
The mass of $\chi_{c1}(nP)$ is selected as $3.901{\rm GeV}$ and $4.178{\rm GeV}$ for $n=2,3$ respectively\cite{Li:2009zu,Beringer:1900zz}.
\label{tab:chic1BESB} }
\centering
\begin{tabular}{c|ccc}
\hline
$\sqrt s$(GeV)& 4.25  & 4.50  & 4.75  \\ \hline
$1P$ OP & $1716\hspace{-0.06cm}\pm\hspace{-0.06cm}466\hspace{-0.06cm}\pm\hspace{-0.06cm}185\hspace{-0.06cm}\pm\hspace{-0.06cm}151\hspace{-0.06cm}\pm\hspace{-0.06cm}86$ & $1127\hspace{-0.06cm}\pm\hspace{-0.06cm}306\hspace{-0.06cm}\pm\hspace{-0.06cm}117\hspace{-0.06cm}\pm\hspace{-0.06cm}64\hspace{-0.06cm}\pm\hspace{-0.06cm}0.3$ & $783\hspace{-0.06cm}\pm\hspace{-0.06cm}213\hspace{-0.06cm}\pm\hspace{-0.06cm}79\hspace{-0.06cm}\pm\hspace{-0.06cm}24\hspace{-0.06cm}\pm\hspace{-0.06cm}27$  \\
$1P$ WP & $1435\hspace{-0.06cm}\pm\hspace{-0.06cm}390\hspace{-0.06cm}\pm\hspace{-0.06cm}156\hspace{-0.06cm}\pm\hspace{-0.06cm}11\hspace{-0.06cm}\pm\hspace{-0.06cm}25$& $967\hspace{-0.06cm}\pm\hspace{-0.06cm}263\hspace{-0.06cm}\pm\hspace{-0.06cm}102\hspace{-0.06cm}\pm\hspace{-0.06cm}16\hspace{-0.06cm}\pm\hspace{-0.06cm}26$& $685\hspace{-0.06cm}\pm\hspace{-0.06cm}186\hspace{-0.06cm}\pm\hspace{-0.06cm}70\hspace{-0.06cm}\pm\hspace{-0.06cm}25\hspace{-0.06cm}\pm\hspace{-0.06cm}39$  \\
\hline
$2P$ OP & $10456\pm3050\pm1099\pm3524$ & $3374\pm984\pm374\pm881$ & $1603\pm468\pm180\pm326$  \\
$2P$ WP & $6809\pm1986\pm700\pm1077$ & $2399\pm700\pm264\pm393$& $1209\pm353\pm136\pm129$  \\
\hline
$3P$ OP  &   & $7586\pm1529\pm784\pm2690$ & $2290\pm462\pm251\pm638$  \\
$3P$ WP  &   & $4831\pm974\pm486\pm1313$& $1597\pm322\pm173\pm292$  \\
\hline\hline
$\sqrt s$(GeV) & 5.00  & 10.6 & 11.2 \\ \hline
$1P$ OP & $568\hspace{-0.06cm}\pm\hspace{-0.06cm}154\hspace{-0.06cm}\pm\hspace{-0.06cm}55\hspace{-0.06cm}\pm\hspace{-0.06cm}5\hspace{-0.06cm}\pm\hspace{-0.06cm}34$ & $15.0\hspace{-0.06cm}\pm\hspace{-0.06cm}4.1\hspace{-0.06cm}\pm\hspace{-0.06cm}0.8\hspace{-0.06cm}\pm\hspace{-0.06cm}2.1\hspace{-0.06cm}\pm\hspace{-0.06cm}2.8$ & $11.9\hspace{-0.06cm}\pm\hspace{-0.06cm}3.2\hspace{-0.06cm}\pm\hspace{-0.06cm}0.6\hspace{-0.06cm}\pm\hspace{-0.06cm}1.8\hspace{-0.06cm}\pm\hspace{-0.06cm}2.3$ \\
$1P$ WP & $505\hspace{-0.06cm}\pm\hspace{-0.06cm}137\hspace{-0.06cm}\pm\hspace{-0.06cm}49\hspace{-0.06cm}\pm\hspace{-0.06cm}26\hspace{-0.06cm}\pm\hspace{-0.06cm}40$ & $14.7\hspace{-0.06cm}\pm\hspace{-0.06cm}4.0\hspace{-0.06cm}\pm\hspace{-0.06cm}0.8\hspace{-0.06cm}\pm\hspace{-0.06cm}2.3\hspace{-0.06cm}\pm\hspace{-0.06cm}2.7$ & $11.7\hspace{-0.06cm}\pm\hspace{-0.06cm}3.2\hspace{-0.06cm}\pm\hspace{-0.06cm}0.6\hspace{-0.06cm}\pm\hspace{-0.06cm}1.9\hspace{-0.06cm}\pm\hspace{-0.06cm}2.3$ \\
\hline
$2P$ OP & $915\pm267\pm102\pm145$ & $10.4\pm3.0\pm0.7\pm1.2$ & $8.2\pm2.4\pm0.5\pm1.0$ \\
$2P$ WP & $720\pm210\pm81\pm47$ & $10.0\pm2.9\pm0.7\pm1.4$ & $7.9\pm2.3\pm0.5\pm1.1$ \\
\hline
$3P$ OP & $1061\pm214\pm119\pm234$ & $7.6\pm1.5\pm0.5\pm0.8$ & $5.9\pm1.2\pm0.4\pm0.6$ \\
$3P$ WP & $786\pm159\pm88\pm97$ & $7.3\pm1.5\pm0.5\pm0.9$ & $5.7\pm1.1\pm0.4\pm0.8$ \\
\hline
\end{tabular}
\end{table*}


\begin{figure}[ht]
\begin{center}
\includegraphics[width=0.86\textwidth]{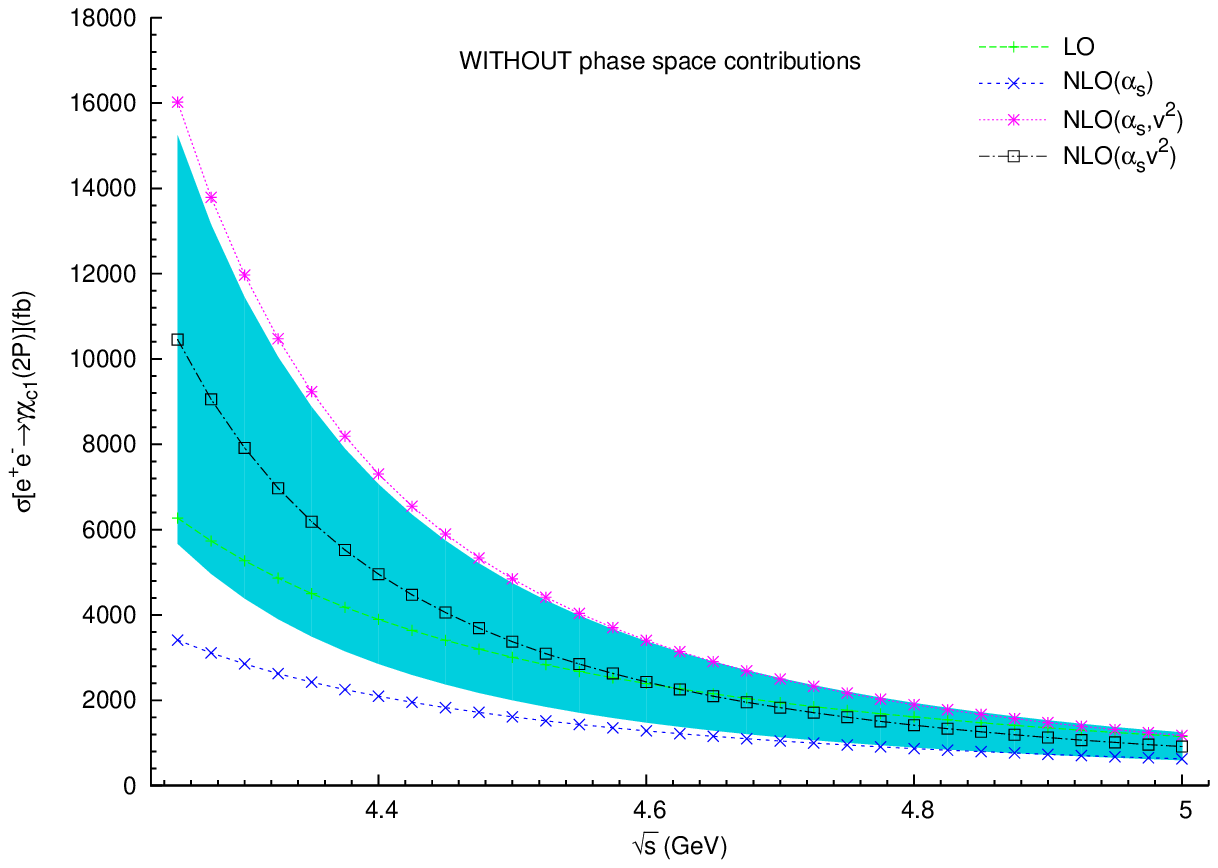}
\includegraphics[width=0.86\textwidth]{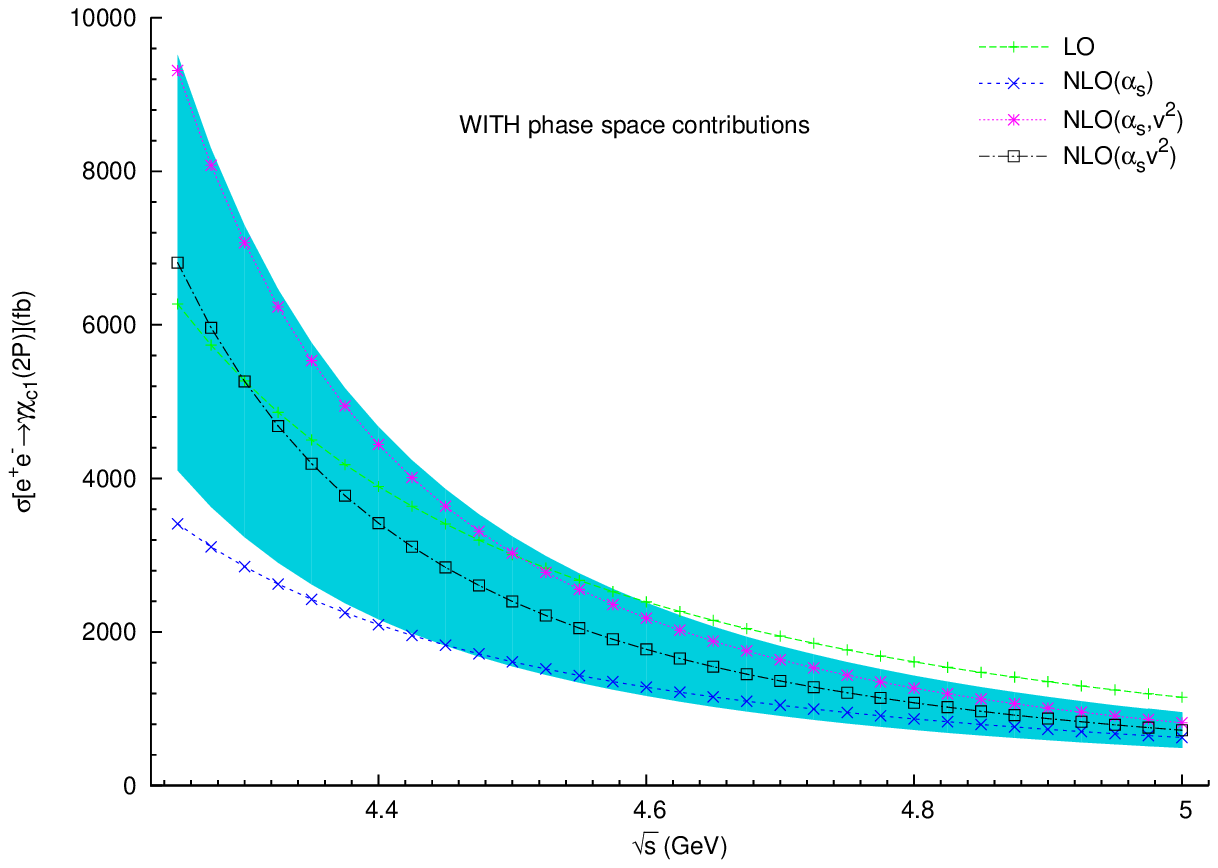}
\end{center}
\caption{\label{fig:chic12p_cs}  The cross sections of the $\chi_{c1}(2P)$ process at the BESIII  energy  region.
The uncertainties for the total cross sections come from the uncertainties of $\as$, $\langle v^2\rangle$ and the wave functions at the origin. }
\end{figure}

\begin{figure}[ht]
\begin{center}
\includegraphics[width=0.86\textwidth]{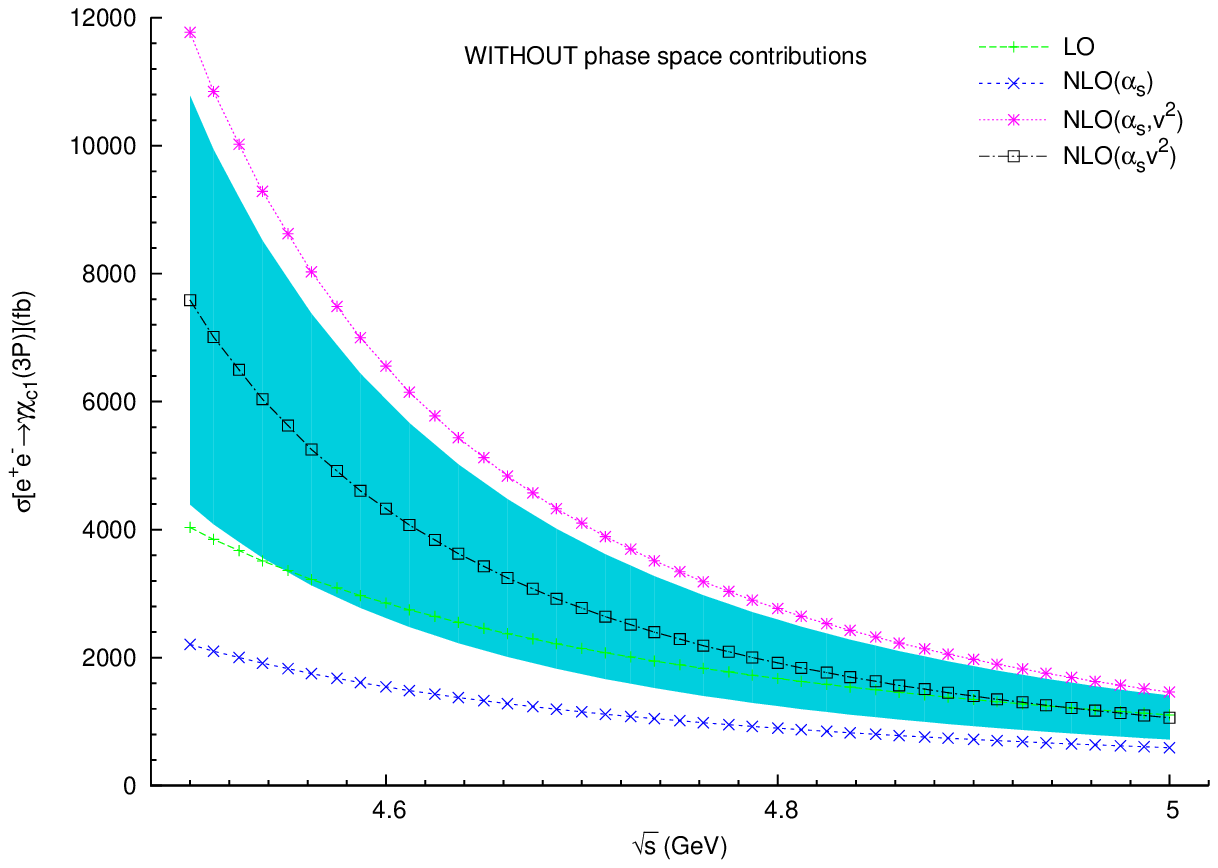}
\includegraphics[width=0.86\textwidth]{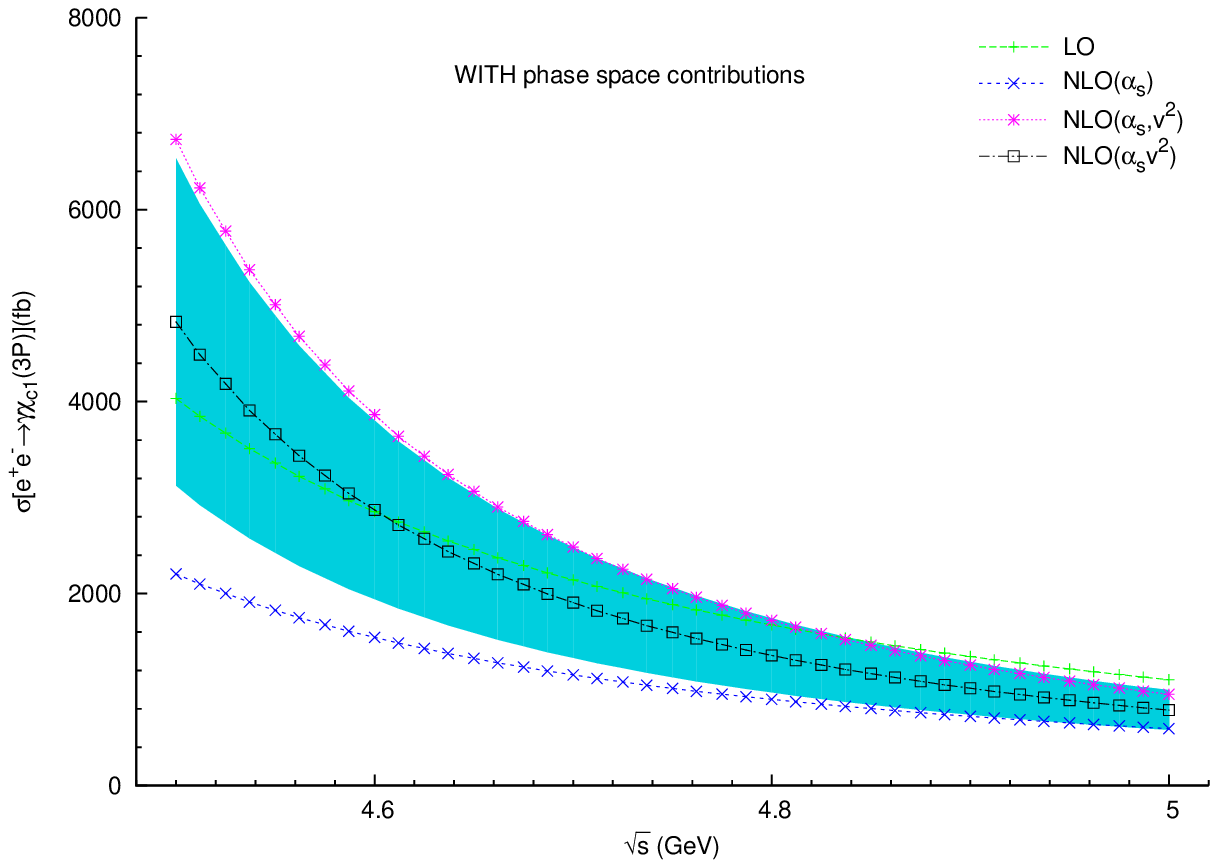}
\end{center}
\caption{\label{fig:chic13p_cs}  The cross sections of the $\chi_{c1}(3P)$ process at the BESIII  energy  region.
The uncertainties for the total cross sections come from the uncertainties of $\as$, $\langle v^2\rangle$ and the wave functions at the origin. }
\end{figure}

\begin{table*}[htbp]\small
\caption{
The total cross sections in ~$fb$~ up to $\as v^2$ order of $e^+e^- \to \chi_{c2}(nP) + \gamma$ with $n=1,2,3$ in the BESIII and B-factories energy region. ¡°WP¡± and ¡°OP¡± indicate considering or ignoring the phase space contributions, respectively.
The uncertainties in each cell come from the uncertainties of the wave functions at the origin, $\as$, $\langle v^2\rangle$, and charm quark mass $m_c$ in turns. For the excited states, we select  the charm quark mass as the half of the meson mass in the calculations, therefore there are no $m_c$ uncertainties.
he mass of $\chi_{c2}(nP)$ is selected as $3.927{\rm GeV}$ and $4.208{\rm GeV}$ for $n=2,3$ respectively\cite{Li:2009zu,Beringer:1900zz}.
\label{tab:chic2BESB} }
\centering
\begin{tabular}{c|ccc}
\hline
$\sqrt s$(GeV)&  4.25  & 4.50  & 4.75 \\ \hline
$1P$ OP &  $1375\hspace{-0.06cm}\pm\hspace{-0.06cm}374\hspace{-0.06cm}\pm\hspace{-0.06cm}211\hspace{-0.06cm}\pm\hspace{-0.06cm}192\hspace{-0.06cm}\pm\hspace{-0.06cm}267$ & $799\hspace{-0.06cm}\pm\hspace{-0.06cm}217\hspace{-0.06cm}\pm\hspace{-0.06cm}128\hspace{-0.06cm}\pm\hspace{-0.06cm}92\hspace{-0.06cm}\pm\hspace{-0.06cm}112$ & $497\hspace{-0.06cm}\pm\hspace{-0.06cm}135\hspace{-0.06cm}\pm\hspace{-0.06cm}82\hspace{-0.06cm}\pm\hspace{-0.06cm}47\hspace{-0.06cm}\pm\hspace{-0.06cm}50$  \\
$1P$ WP &  $1178\hspace{-0.06cm}\pm\hspace{-0.06cm}320\hspace{-0.06cm}\pm\hspace{-0.06cm}179\hspace{-0.06cm}\pm\hspace{-0.06cm}94\hspace{-0.06cm}\pm\hspace{-0.06cm}193$& $701\hspace{-0.06cm}\pm\hspace{-0.06cm}191\hspace{-0.06cm}\pm\hspace{-0.06cm}111\hspace{-0.06cm}\pm\hspace{-0.06cm}43\hspace{-0.06cm}\pm\hspace{-0.06cm}82$& $443\hspace{-0.06cm}\pm\hspace{-0.06cm}121\hspace{-0.06cm}\pm\hspace{-0.06cm}73\hspace{-0.06cm}\pm\hspace{-0.06cm}21\hspace{-0.06cm}\pm\hspace{-0.06cm}36$  \\
\hline
$2P$ OP   & $17037\pm4969\pm1898\pm6041$ & $4564\pm1331\pm568\pm1305$ & $1878\pm548\pm252\pm444$  \\
$2P$ WP   & $11250\pm3281\pm1205\pm3147$ & $3316\pm967\pm402\pm681$& $1451\pm423\pm190\pm230$  \\
\hline
$3P$ OP  &   & $13164\pm2654\pm1424\pm4895$ & $3253\pm656\pm395\pm983$  \\
$3P$ WP   &  & $8465\pm1707\pm878\pm2546$& $2314\pm466\pm273\pm513$  \\
\hline\hline
$\sqrt s$(GeV)&   5.00  & 10.6 & 11.2 \\ \hline
$1P$ OP & $325\hspace{-0.06cm}\pm\hspace{-0.06cm}88\hspace{-0.06cm}\pm\hspace{-0.06cm}55\hspace{-0.06cm}\pm\hspace{-0.06cm}25\hspace{-0.06cm}\pm\hspace{-0.06cm}23$ & $3.1\hspace{-0.06cm}\pm\hspace{-0.06cm}0.8\hspace{-0.06cm}\pm\hspace{-0.06cm}0.8\hspace{-0.06cm}\pm\hspace{-0.06cm}0.2\hspace{-0.06cm}\pm\hspace{-0.06cm}0.4$ & $2.3\hspace{-0.06cm}\pm\hspace{-0.06cm}0.6\hspace{-0.06cm}\pm\hspace{-0.06cm}0.6\hspace{-0.06cm}\pm\hspace{-0.06cm}0.2\hspace{-0.06cm}\pm\hspace{-0.06cm}0.3$ \\
$1P$ WP & $294\hspace{-0.06cm}\pm\hspace{-0.06cm}80\hspace{-0.06cm}\pm\hspace{-0.06cm}50\hspace{-0.06cm}\pm\hspace{-0.06cm}10\hspace{-0.06cm}\pm\hspace{-0.06cm}16$ & $3.0\hspace{-0.06cm}\pm\hspace{-0.06cm}0.8\hspace{-0.06cm}\pm\hspace{-0.06cm}0.8\hspace{-0.06cm}\pm\hspace{-0.06cm}0.2\hspace{-0.06cm}\pm\hspace{-0.06cm}0.4$ & $2.3\hspace{-0.06cm}\pm\hspace{-0.06cm}0.6\hspace{-0.06cm}\pm\hspace{-0.06cm}0.6\hspace{-0.06cm}\pm\hspace{-0.06cm}0.2\hspace{-0.06cm}\pm\hspace{-0.06cm}0.3$ \\
\hline
$2P$ OP   & $945\pm275\pm133\pm188$ & $2.7\pm0.8\pm0.6\pm0.1$ & $2.0\pm0.6\pm0.5\pm0.1$ \\
$2P$ WP   & $761\pm222\pm106\pm96$ & $2.6\pm0.8\pm0.6\pm0.1$ & $1.9\pm0.6\pm0.5\pm0.1$ \\
\hline
$3P$ OP  & $1305\pm263\pm171\pm328$ & $2.2\pm0.4\pm0.5\pm0.$ & $1.6\pm0.3\pm0.4\pm0.$ \\
$3P$ WP  & $990\pm200\pm127\pm171$ & $2.1\pm0.4\pm0.5\pm0.1$ & $1.5\pm0.3\pm0.3\pm0.1$ \\
\hline
\end{tabular}
\end{table*}


\begin{figure}[ht]
\begin{center}
\includegraphics[width=0.86\textwidth]{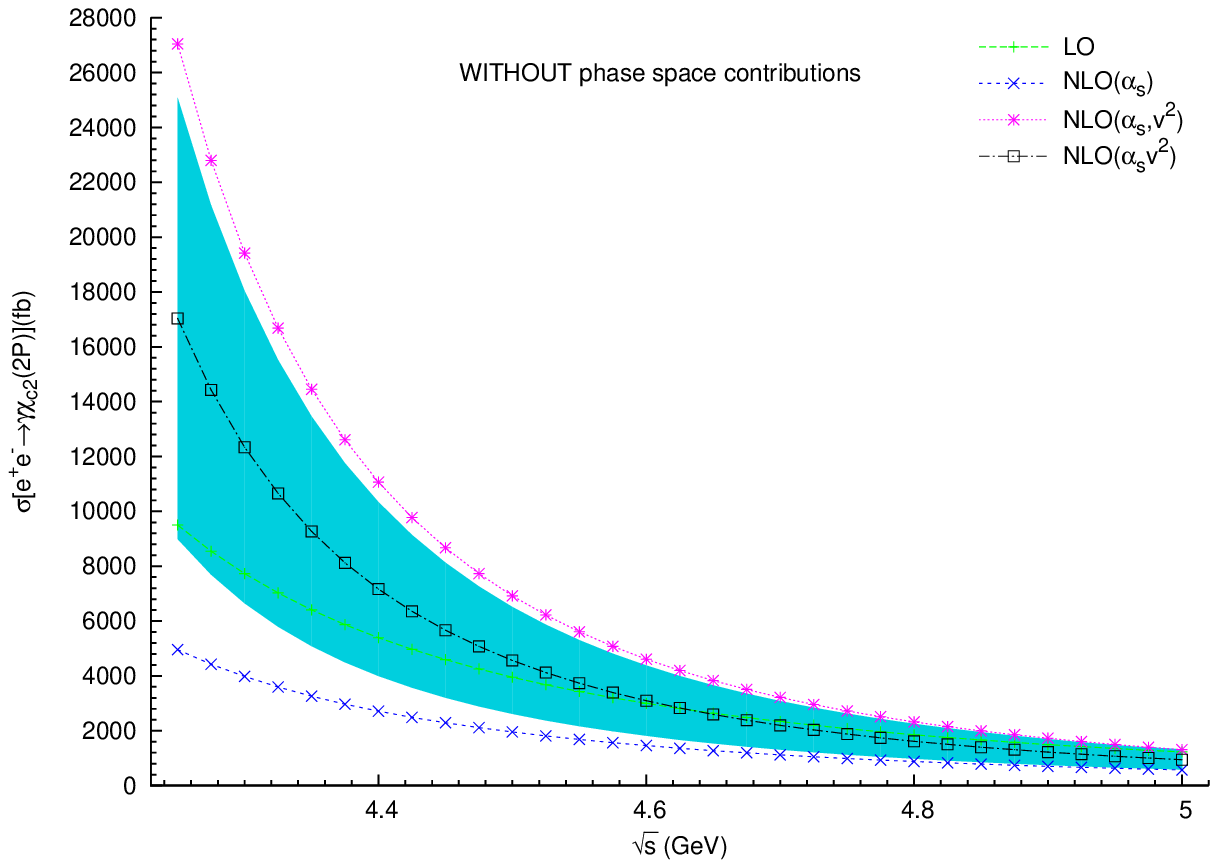}
\includegraphics[width=0.86\textwidth]{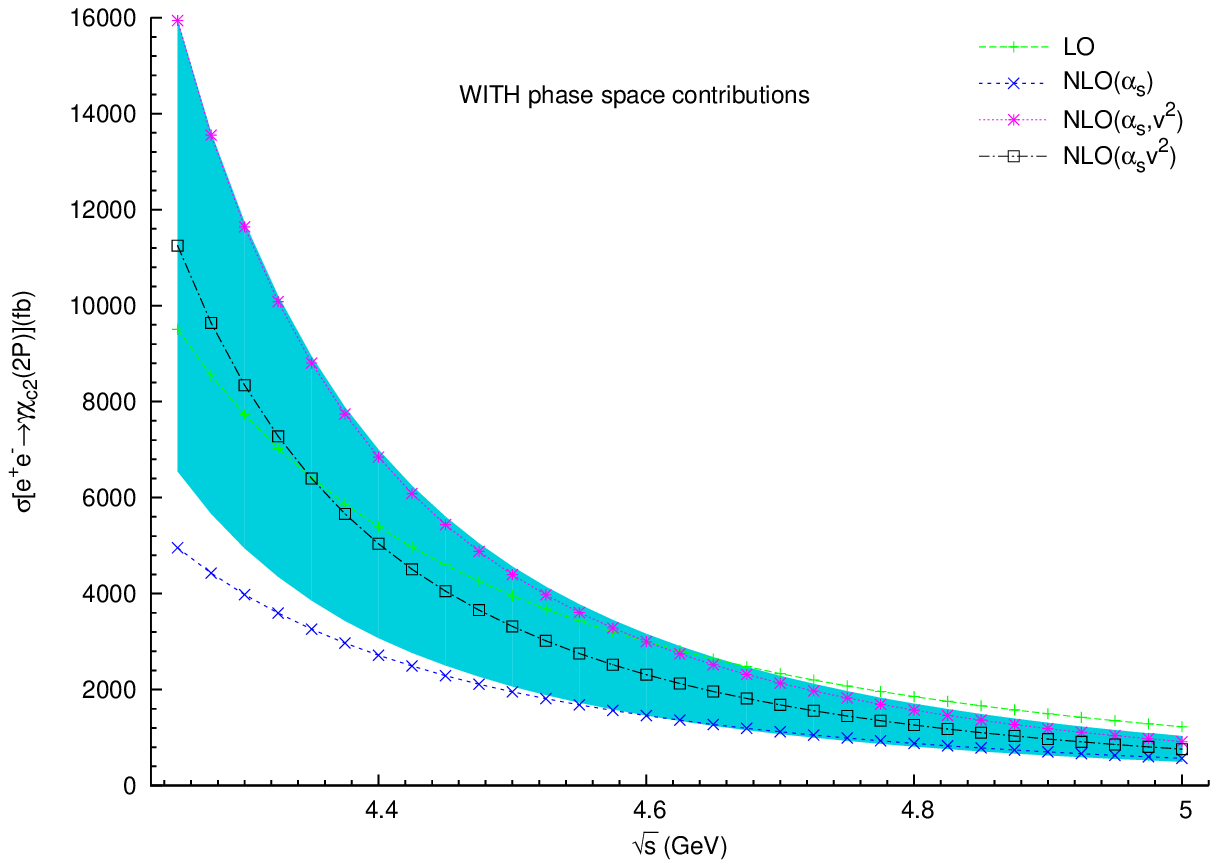}
\end{center}
\caption{\label{fig:chic22p_cs}  The cross sections of the $\chi_{c2}(2P)$ process at the BESIII  energy  region.
The uncertainties for the total cross sections come from the uncertainties of $\as$, $\langle v^2\rangle$ and the wave functions at the origin. }
\end{figure}

\begin{figure}[ht]
\begin{center}
\includegraphics[width=0.86\textwidth]{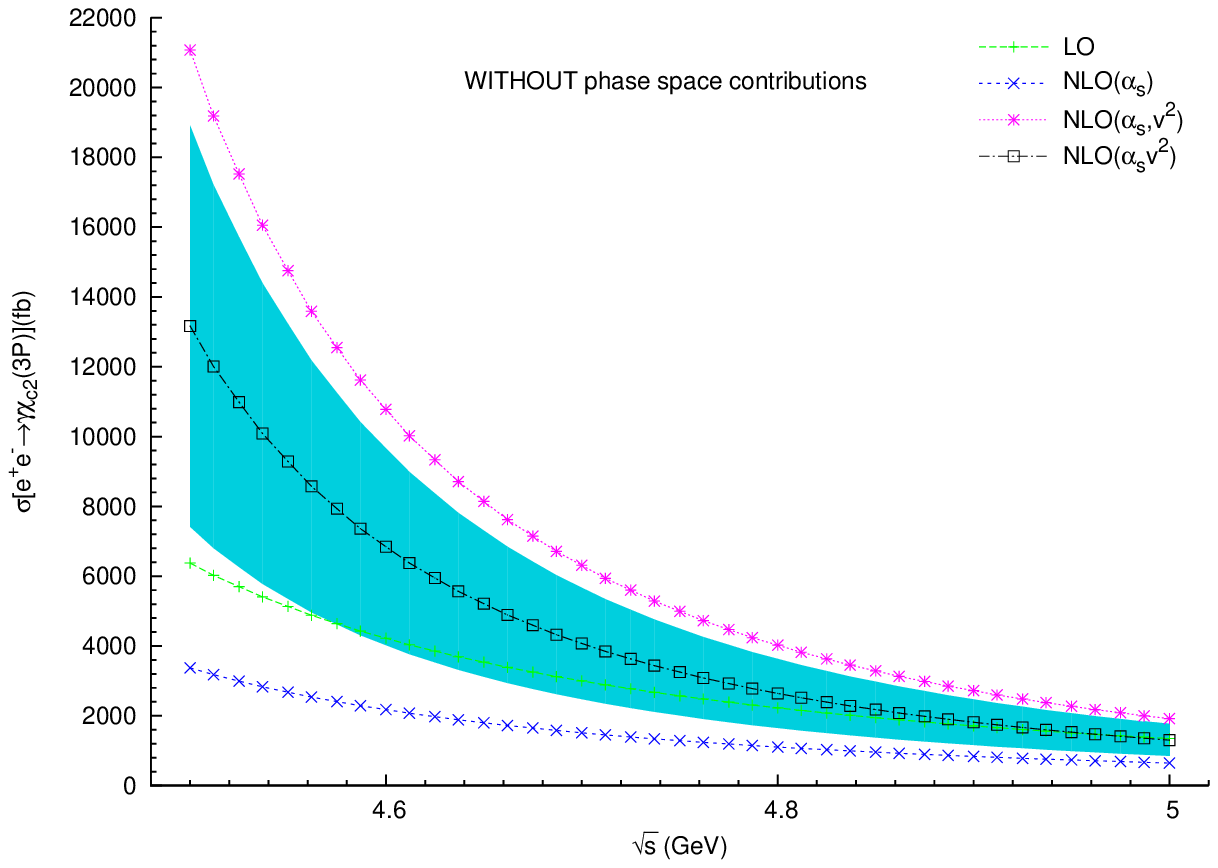}
\includegraphics[width=0.86\textwidth]{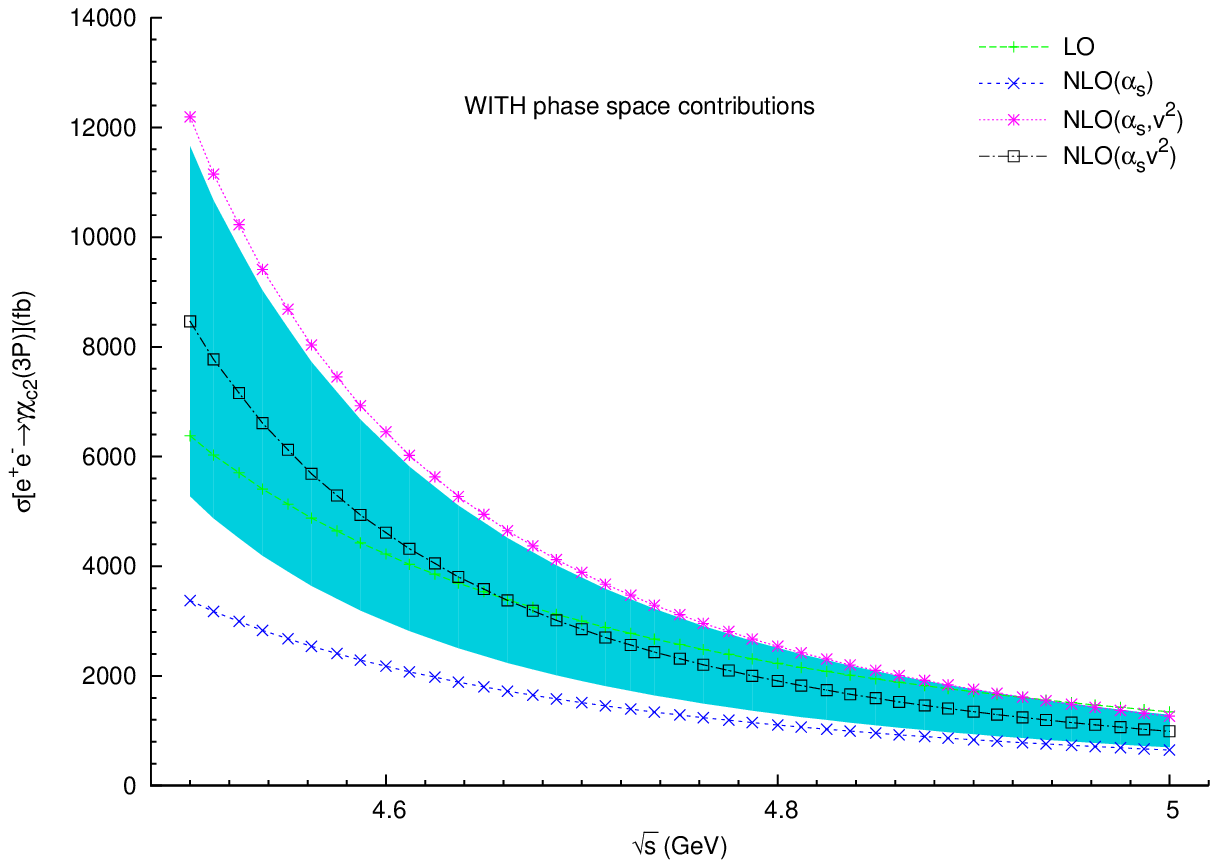}
\end{center}
\caption{\label{fig:chic23p_cs}  The cross sections of the $\chi_{c2}(3P)$ process at the BESIII  energy  region.
The uncertainties for the total cross sections come from the uncertainties of $\as$, $\langle v^2\rangle$ and the wave functions at the origin. }
\end{figure}

For the high $\eta_c(ns)$ and  $\chi_{cJ}(nP)$  states, the masses of these states extremely approximate the BESIII beam energy. NRQCD factorization will be broken down near the endpoint.
In our previous works by Ref.\cite{Li:2013nna}, the  charm quark mass is set the half of the meson.
But in Ref\cite{Li:2009ki,Chao:2013cca}, different strategy is used to remedy the phase space integrand near the threshold, an additional unitary factor is introduced, and the charm quark mass is set about $1.5~$GeV. Unfortunately, they obtain significantly different cross sections for the production of these near-threshold particles for the excited $P$-wave states.
We remain the strategy in our previous work to set the quark mass to the half of the meson.
The results are shown in Tab.\ref{tab:etaBESB}, \ref{tab:chic0BESB}, \ref{tab:chic1BESB} \ref{tab:chic2BESB}  and in Fig.\ref{fig:chic12p_cs}, \ref{fig:chic13p_cs}, \ref{fig:chic22p_cs}, \ref{fig:chic23p_cs}.
The numerical cross sections for $\eta_c(2S)$ states positively increase compared with those for $\mathcal{O}(\as+v^2)$. Meanwhile, the cross-sections for  the excited $P$-wave states are lower than those from the previous $\mathcal{O}(\as+v^2)$ results.  However for $\eta_c(3S)$ state, the numerical values are still assigned to BESIII to determine the states.
For excited $P$-wave states, the cross sections come down compared with the previous $\mathcal{O}(\as+v^2)$ results.
But the numerical values are still referred for BESIII to find these states.

\begin{figure}[ht]
\begin{center}
\includegraphics[width=0.86\textwidth]{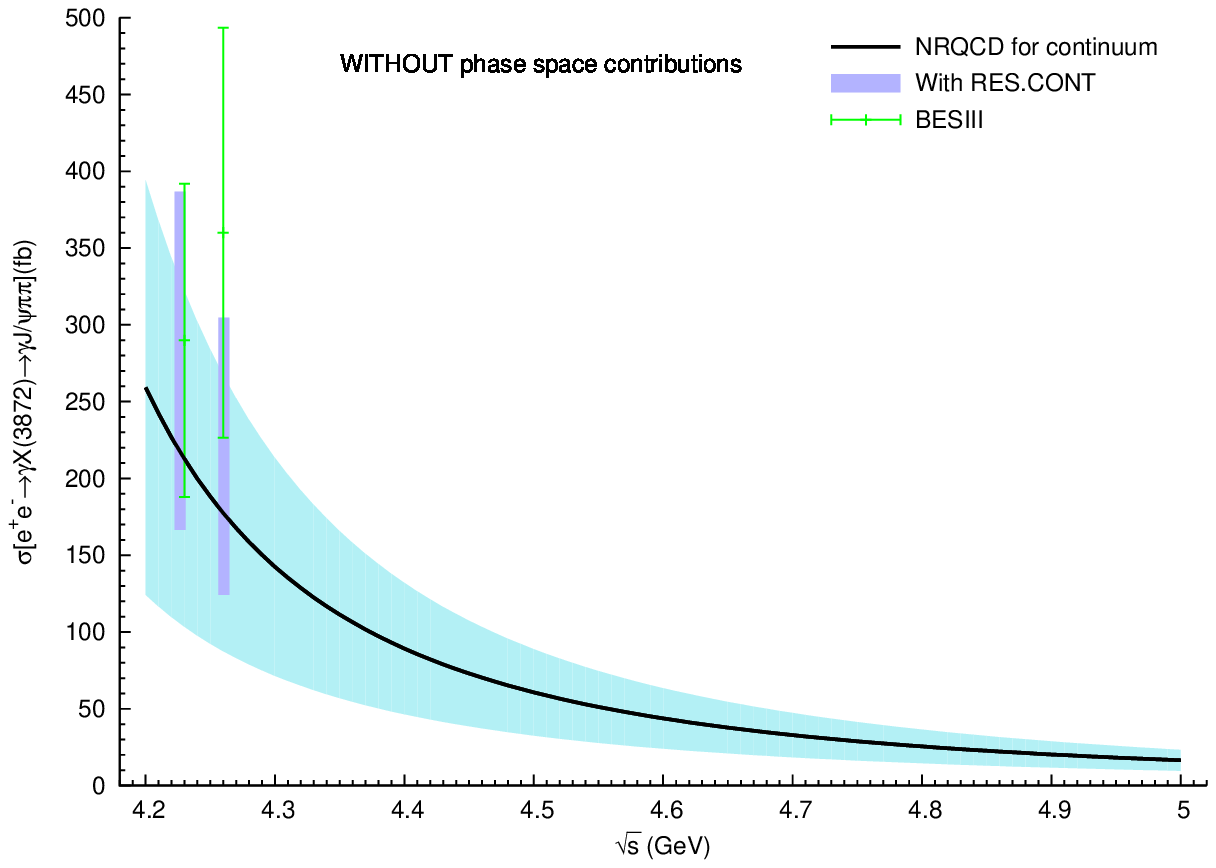}
\includegraphics[width=0.86\textwidth]{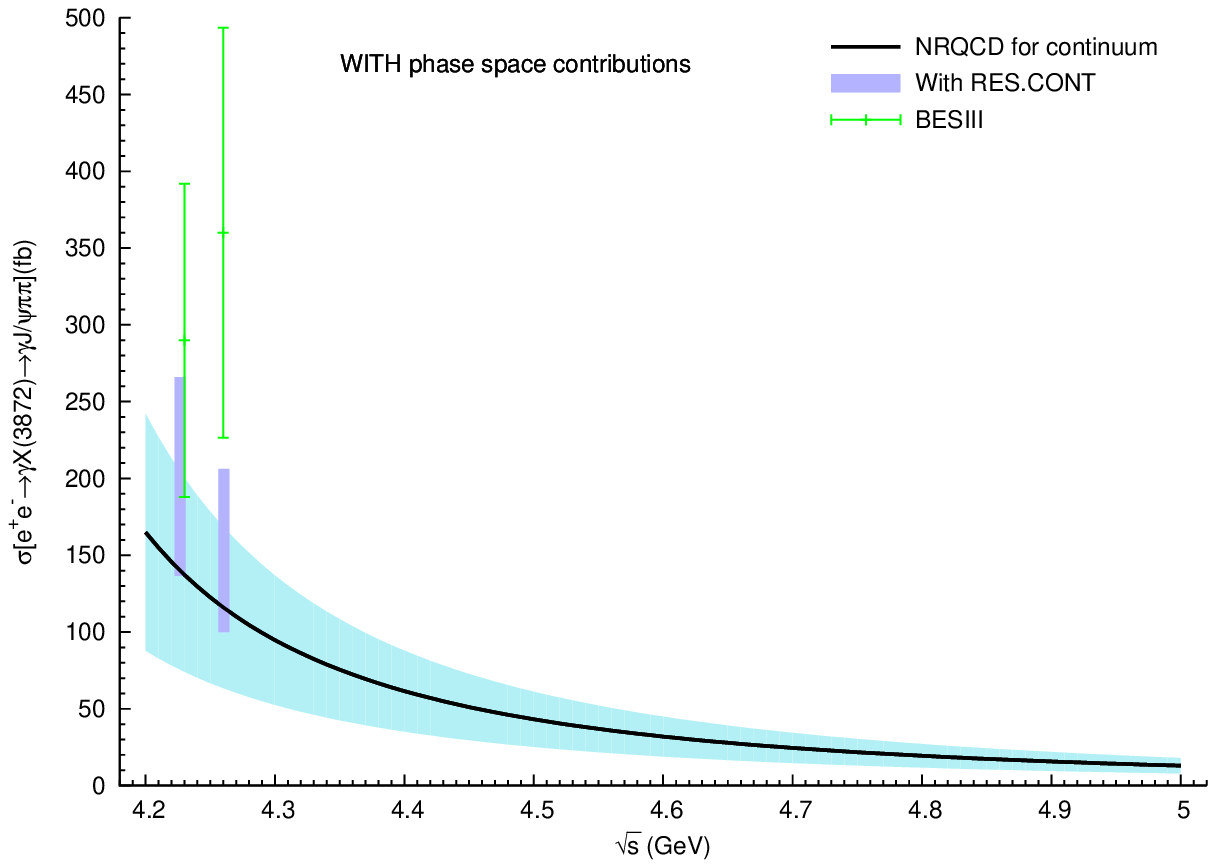}
\end{center}
\caption{\label{fig:x3872_cs}  The cross sections of the $X(3872)$ process at the BESIII energy region when taking $X(3872)$ as the mixture with $\chi_{c1}(2P)$ component.
The uncertainties for the total cross sections come from the uncertainties of $m_c$, $\as$, $\langle v^2\rangle$, and the wave functions at the origin. "With RES.CONT" means considering the contributions from both continuum and resonance.}
\end{figure}

As discussed in our previous works, the results of $\eta_c(mS)$ and $\chi_{cJ}(nP)$ states are helpfull charifly the nature of $XYZ$ particles with the even charge conjugation, such as $X(3872)$, $X(3940)$, $X(4160)$ and $X(4350)$.
Taking $X(3872)$ state for an example, we considered it as the mixture with $\chi_{c1}(2P)$ component\cite{Li:2013nna}, therefore, the cross sections for $X(3872)$ are determined by
\bqn
d\sigma[\ee\rightarrow\gamma X(3872){\rightarrow}\gamma J/\psi\pi^+\pi^-]=d\sigma[\ee\rightarrow\gamma \chi_{c1}(2P)]{\times}k,
\eqn
where $k=Z^{X(3872)}_{c\bar{c}}{\times }Br[X(3872) \to J/\psi \pi^+\pi^-]$.
$Br[X(3872) \to J/\psi \pi^+\pi^-]$ is the branching fraction for $X(3872)$
decay to $J/\psi \pi^+\pi^-$. $Z^{X(3872)}_{c\bar{c}}$ is the probability of the
$\chi_{c1}(2P)$ component in $X(3872)$. $k=0.018 \pm 0.04 $ \cite{Meng:2005er,Meng:2013gga}. With the results up-to $\mathcal{O}(\as v^2)$, we revisit the cross sections for $X(3872)$ shown in Fig.\ref{fig:x3872_cs}.
In the figure, we also give the total cross sections at the data points for the BESIII measurements including the contributions from the resonances ($\psi(4040)$ and $\psi(4160)$) which have been discussed in our previous paper and are listed here
\bqna
&&(\sigma_{\psi(4040)}[4.23]+\sigma_{\psi(4160)}[4.23])\times k= (62 \pm 14) fb,\nonumber \\
&&(\sigma_{\psi(4040)}[4.26]+\sigma_{\psi(4160)}[4.26])\times k= (37 \pm 8) fb.
\eqna
From the figure, the cross sections for the predictions of $X(3872)$ may be smaller than the experiment data, but one still can not jump to conclusions for the nature of the $X(3872)$ and the more data are required.

\section{Summary}~\label{sec_summary}

In this study, we extend our previous works on the production of charmonia with even charge conjugation  in the processes $e^+e^-\to\eta_c(nS)(\chi_{cJ}(mP)) + \gamma$ up to the $\mathcal{O}(\as v^2)$ corrections. The results indicate that these corrections exhibit a logarithmic singularity of $\ln(1-r)$, which is not observed in the $\mathcal{O}(\as)$ corrections near the threshold. The $\mathcal{O}(\as v^2)$ corrections also contribute to the total cross-sections near the threshold and are important to the di-photon decay for the $\chi_{c0}$ and $\chi_{c2}$ states. We revisit the numerical calculations to the cross -sections for the $\eta_c(nS)$ and $\chi_{cJ}(mP)$ states using the results for the $\mathcal{O}(\as v^2)$ corrections.

\acknowledgments{
The authors would like to thank Professor C.P. Shen for useful
discussion. This work was
supported by the National Natural Science
Foundation of China (Grants No. 11375021), the Foundation for the Author of National
Excellent Doctoral Dissertation of China (Grants No. 2007B18 and No. 201020), the New Century Excellent Talents in University (NCET) under grant
NCET-13-0030,  the Major State Basic Research Development Program of China (No. 2015CB856701), and the Education Ministry of
LiaoNing Province.}

\section{Appendix}

In this section, we give the matching results of the short-distance coefficients for $\eta_c$ process.
The Lorentz invariance determines the amplitude should have the form of $$A\epsilon^{\mu\nu\rho\tau}(\eps^{\ast}_{Q})_{\mu}(\eps^{\ast}_{k})_{\nu}k_{\rho}p_{\tau}.$$
Therefore the coefficients in $v^{(0)}$ must be like $A^{(0)}\eps_1$.
The $\mathcal{O}(v^2)$ coefficients obtained in proceed of derivate the amplitude will be like $A^{(v^2)}\eps_1+B^{(v^2)}\eps_2$.
$\eps_1$ and $\eps_2$ have defined as Eq.(\ref{eq_e1e2def}).
Therefore we write the $\mathcal{O}(v^2)$ short-distance coefficients into a plus of three parts as seen in the following results.
The third term is just to cancel the $\mathcal{O}(v^2)$ contributions from the relativistic normalization factor in Eq.(\ref{eq_nrqcd_amp_ex_asv2}). And we omit the imaginary parts in the coefficients in $\mathcal{O}(\as)$ and $\mathcal{O}(\as v^2)$, which do't contribute to cross sections at order of $\mathcal{O}(\as v^2)$.

\begin{eqnarray}
d^{(0)}{\equiv}A^{(0)}\eps_1=\frac{(4\pi\alpha)Q_c^2 r}{2m_c^3 (1-r)}\eps_1.
\end{eqnarray}
\begin{eqnarray}
d^{(v^2)}=-\frac{(4 \pi  \alpha)  Q_c^2 r (5 -17 r)}{24 m_c^5 (1-r)^2}\eps_1
   +\frac{A^{(0)}}{m_c^2}\eps_2
   -\frac{d^{(0)}}{4m_c^2}.
\end{eqnarray}

\begin{eqnarray}
&& d^{(\as)}{\equiv}A^{(\as)}\eps_1=\frac{(4\pi\alpha) (4\pi\as)
   Q_c^2 C_A C_F r}{96
   \pi ^2 m_c^3 N_c
   (2-r)^2 (1-r)^2}\eps_1\nonumber\\
   &&\big\{
   -6 (1-r) [5 r^2-r
   (19+\ln 16)+18+\ln
   64]-\pi ^2 (1+r)
   (2-r)^2+6 [r
   (r+2)-6] \ln r
   \nonumber\\
   &&+ 18(2-r)^2[(1-\sqrt{1-r})\ln
   (1-\sqrt{1-r}) + (1+\sqrt{1-r})\ln
   (1+\sqrt{1-r})] \nonumber\\
   &&-12 (1-r) (3- 2 r)\ln (1-r)
   +3 (2-r)^2[r \text{Li}_2(\frac{2}{r}-1)
   + (2+r)\big(\text{Li}_2(\frac{2}{1-\sqrt{1-r}})\nonumber\\
   &&+\text{Li}_2(\frac{2}{1+\sqrt{1-r}})+\text{Li}_2(\frac{r}{2-r})-\text{Li}_2(\frac{2}{2-r})-\text{Li}_2(\frac{2}{r})\big)
   \big\}.
\end{eqnarray}

\begin{eqnarray}
&& d^{(\as v^2)}=  -\frac{(4\pi\alpha) (4\pi\as) Q_c^2 C_A C_F r
   }{1152 \pi
   ^2 m_c^5 N_c
   (2-r)^4 (1-r)^3}\eps_1\nonumber\\
   &&\big\{ -2(1-r)[r^5 (96
   \ln2-211)-156r^4 (3\ln2-10)+23 r^3 (30 \ln2 -181) \nonumber\\ &&+r^2 (4598-606 \ln2) +36 r (24\ln2 - 37)-8
   (79+87 \ln2)]
   +\pi ^2 (21r^2+28r-5) (2-r)^4\nonumber\\
   &&-6  (32 r^6-251 r^5+889
   r^4-1936 r^3+2482 r^2-1496
   r+216)\ln r
   \nonumber\\
   &&-6(2-r)^4 (63 r+1)[(1-\sqrt{1-r})\ln
   (1-\sqrt{1-r}) + (1+\sqrt{1-r})\ln
   (1+\sqrt{1-r})]\nonumber\\
   &&-12 (1-r) (16 r^5-78 r^4+115 r^3-101 r^2+144 r-116)\ln(1-r)\nonumber\\
   &&
   -3 (2-r)^4[3r(1+7r) \text{Li}_2(\frac{2}{r}-1)
   + (21r^2+53r-10)\big(\text{Li}_2(\frac{2}{1-\sqrt{1-r}})\nonumber\\
   &&+\text{Li}_2(\frac{2}{1+\sqrt{1-r}})+\text{Li}_2(\frac{r}{2-r})-\text{Li}_2(\frac{2}{2-r})-\text{Li}_2(\frac{2}{r})\big)
   \big\}
   +\frac{A^{(\as)}}{m_c^2}\eps_2-\frac{d^{(\as)}}{4m_c^2}.
\end{eqnarray}


\providecommand{\href}[2]{#2}\begingroup\raggedright\endgroup
\end{document}